\documentclass[a4paper,11pt]{article}

\usepackage{jcappub} 

\title{\Large{Schwinger-Keldysh mechanism in extended quasi single field inflation}}

\author[a,b]{Haidar Sheikhahmadi}

\affiliation[a]{School of Astronomy, Institute for Research in
Fundamental Sciences (IPM), P. O. Box 19395-5531, Tehran, Iran}
\affiliation[b]{Center for Space Research, North-West University, Mafikeng, South Africa}
\emailAdd{h.sh.ahmadi@gmail.com;h.sheikhahmadi@ipm.ir}

\abstract{We study an extension of quasi single field model of inflation containing multiple semi-heavy isocurvaton fields using the Schwinger-Keldysh mechanism. We calculate the amplitudes and the shapes of the  bispectrum and the trispectrum. We show that the diagrammatic approach associated with the   Schwinger-Keldysh mechanism simplifies the analysis considerably compared to standard in-in formalism.  This method is helpful to study the spectroscopy of masses and couplings of light and semi-heavy fields during inflation.\\
\\
\textbf{pacs}:~04.20.-q, 04.20.CV, 04.20.Dw, 04.25.dc\\
\\
\textit{keywords:} Diagrammatic Schwinger-Keldysh rules,  In-in formalism, Quasi-single-multifields scenario, Non-Gaussianities}
\usepackage{mathrsfs}
\begin{document}
\maketitle
\flushbottom
\newpage

\renewcommand{\arraystretch}{1.2}
\def\balpha{\mbox{\boldmath $\alpha$}}
\def\bsigma{\mbox{\boldmath $\sigma$}}
\def\balpha{\mbox{\boldmath $\alpha$}}
\def\bsigma{\mbox{\boldmath $\sigma$}}
\def\bC{{\bf C}}
\def\bc{{\bf c}}
\def\bk{{\bf k}}
\def\bp{{\bf p}}
\def\bq{{\bf q}}
\def\bx{{\bf x}}
\def\by{{\bf y}}
\def\ve{{\varepsilon}}
\def\Mpl{M_{\rm P}}
\def\mpl{M_{\rm P}}
\def\half{\frac{1}{2}}
\def\gev{{~{\rm GeV}}}
\newcommand{\ba}{\begin{eqnarray}}
\newcommand{\ea}{\end{eqnarray}}

\def\bC{{\bf C}}
\def\bc{{\bf c}}
\def\bk{{\bf k}}
\def\bp{{\bf p}}
\def\bq{{\bf q}}
\def\bx{{\bf x}}
\def\by{{\bf y}}
\def\ve{{\varepsilon}}

\def\CA{{\cal A}}
\def\CE{{\cal E}}
\def\CH{{\cal H}}
\def\CL{{\cal L}}
\def\CO{{\cal O}}
\def\CP{{\cal P}}
\def\CV{{\cal V}}

\def\Ci{{\rm Ci}}
\def\Si{{\rm Si}}

\def\ty{{\tilde y}}
\def\ttau{{\tilde \tau}}
\def\txi{{\tilde\xi}}

\def\cPR{{\cal P}_{\cal R}}

\def\high{\vphantom{\Biggl(}\displaystyle}

\def\Mpl{M_{\rm P}}
\def\mpl{M_{\rm P}}
\def\half{\frac{1}{2}}
\def\gev{{~{\rm GeV}}}
\def\bge{\begin{equation}}
\def\ede{\end{equation}}
\def\bga{\begin{aligned}}
\def\eda{\end{aligned}}
\def\bgp{\begin{pmatrix}}
\def\edp{\end{pmatrix}}
\def\bgs{\begin{subequations}}
\def\eds{\end{subequations}}

\def\di{{\mathrm{d}}}
\def\D{{\mathrm{D}}}
\def\Di{{\mathcal{D}}}

\newcommand{\FR}[2]{\displaystyle\frac{\,{#1}\,}{#2}}
\newcommand{\fr}[2]{\mbox{$\frac{\,{#1}\,}{#2}$}}
\newcommand{\n}{\nonumber}
\renewcommand{\rm}{\mathrm}
\renewcommand{\thefootnote}{\arabic{footnote}}
 \graphicspath{{fig/}}

\newcommand{\red}{\color{red}}
\newcommand{\blue}{\color{blue}}
\newcommand{\gray}{\color{gray}}
\newcommand{\hl}{\color[rgb]{1,0,.5}}

\def\bge{\begin{equation}}
\def\ede{\end{equation}}
\def\bga{\begin{aligned}}
\def\eda{\end{aligned}}
\def\bgp{\begin{pmatrix}}
\def\edp{\end{pmatrix}}
\def\bgs{\begin{subequations}}
\def\eds{\end{subequations}}
\def\di{{\mathrm{d}}}
\def\D{{\mathrm{D}}}
\def\Di{{\mathcal{D}}}

\def\T{\mathcal{T}}
\def\mb{\mathbf}
\def\ms{\mathscr}
\def\mf{\mathfrak}
\def\mc{\mathcal}
\def\cl{\mathrm{cl}}
\def\pd{\partial}
\def\ld{{\mathscr{L}}}
\def\hd{{\mathscr{H}}}
\def\z{{\bar{z}}}
\def\la{\langle}\def\ra{\rangle}
\def\sla{\slashed}
\def\const{\mathrm{const.}}
\setlength\unitlength{1mm}
\def\tr{\mathrm{\,tr\,}}
\def\Tr{\mathrm{\,Tr\,}}
\def\Det{\mathrm{\,Det\,}}
\def\to{\rightarrow}
\def\To{\Rightarrow}
\def\ii{\mathrm{i}}

\def\al{\alpha}
\def\be{\beta}
\def\ga{\gamma}
\def\de{\delta}
\def\ep{\epsilon}
\def\ka{\kappa}
\def\lam{\lambda}
\def\rh{\rho}
\def\si{\sigma}
\def\ze{\zeta}

\def\C{\mathbb{C}}
\def\R{\mathbb{R}}

\def\mk{{\mb{k}}}
\def\mx{{\mb{x}}}

\def\Mp{M_{\text{Pl}}}

\def\m{{\mathrm{M}}}

\def\Re{\mathrm{Re}\,}
\def\Im{\mathrm{Im}\,}

\def\ad{\mathrm{\,ad\,}}

\def\LIR{\Lambda_{\text{IR}}}

\def\eff{{\mathrm{eff}}}

\def\C{\mathbb{C}}
\def\R{\mathbb{R}}

\def\Mp{M_{\text{Pl}}}

\def\Re{\mathrm{Re}\,}
\def\Im{\mathrm{Im}\,}

\def\ad{\mathrm{\,ad\,}}

\def\T{\mathcal{T}}
\def\mb{\mathbf}
\def\ms{\mathscr}
\def\mf{\mathfrak}
\def\mc{\mathcal}
\def\cl{\mathrm{cl}}
\def\pd{\partial}
\def\ld{{\mathscr{L}}}
\def\hd{{\mathscr{H}}}
\def\z{{\bar{z}}}
\def\la{\langle}\def\ra{\rangle}
\def\sla{\slashed}
\def\const{\mathrm{const.}}
\setlength\unitlength{1mm}
\def\tr{\mathrm{\,tr\,}}
\def\Tr{\mathrm{\,Tr\,}}
\def\Det{\mathrm{\,Det\,}}
\def\to{\rightarrow}
\def\To{\Rightarrow}
\def\ii{\mathrm{i}}

\def\al{\alpha}
\def\be{\beta}
\def\ga{\gamma}
\def\de{\delta}
\def\ep{\epsilon}
\def\ka{\kappa}
\def\lam{\lambda}
\def\rh{\rho}
\def\si{\sigma}
\def\ze{\zeta}

\def\C{\mathbb{C}}
\def\R{\mathbb{R}}

\def\mk{{\mb{k}}}
\def\mx{{\mb{x}}}

\def\Mp{M_{\text{Pl}}}

\def\m{{\mathrm{M}}}

\def\Re{\mathrm{Re}\,}
\def\Im{\mathrm{Im}\,}

\def\ad{\mathrm{\,ad\,}}

\def\LIR{\Lambda_{\text{IR}}}

\def\eff{{\mathrm{eff}}}

\newcommand{\ob}[1]{\mkern 2mu \overline{\mkern -2mu #1 \mkern -2mu}\mkern 2mu}
\newcommand{\wt}[1]{\mkern 2mu \widetilde{\mkern -2mu #1 \mkern -2mu}\mkern 2mu}
\newcommand{\wh}[1]{\mkern 2mu \widehat{\mkern-2mu#1\mkern-2mu}\mkern 2mu}

\newcommand{\bea}{\begin{eqnarray}}
\newcommand{\eea}{\end{eqnarray}}
\newcommand{\barr}{\begin{array}}
\newcommand{\earr}{\end{array}}

\section{Convention}
Herewith, briefly, we supply the explanation for diagrams and some rules due to them.

\begin{figure}[ht]
\centering
\includegraphics[scale=.950]{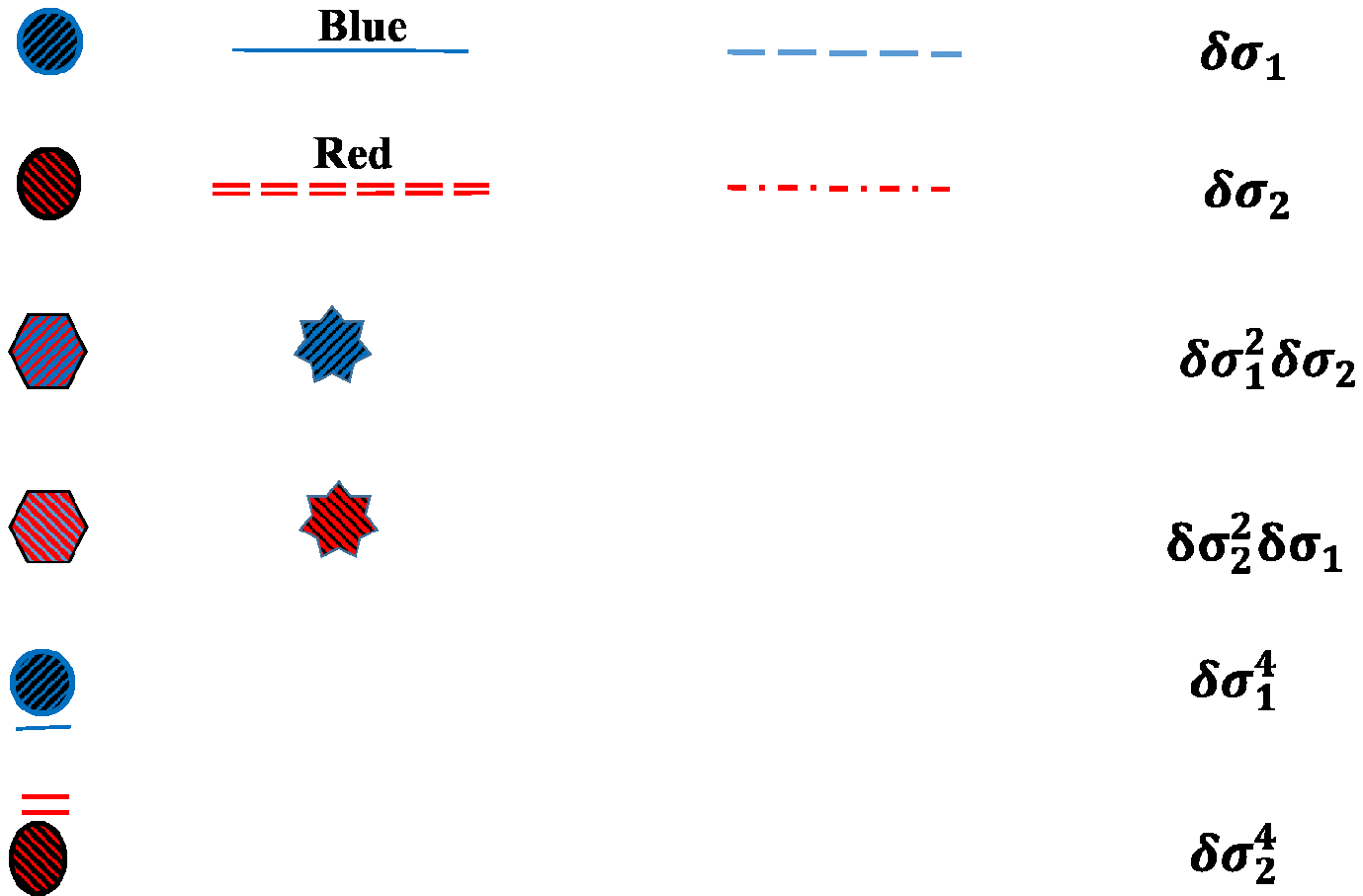}
\label{diagthreepointmain}
\end{figure}
\begin{figure}[ht]
\centering
\includegraphics[scale=.950]{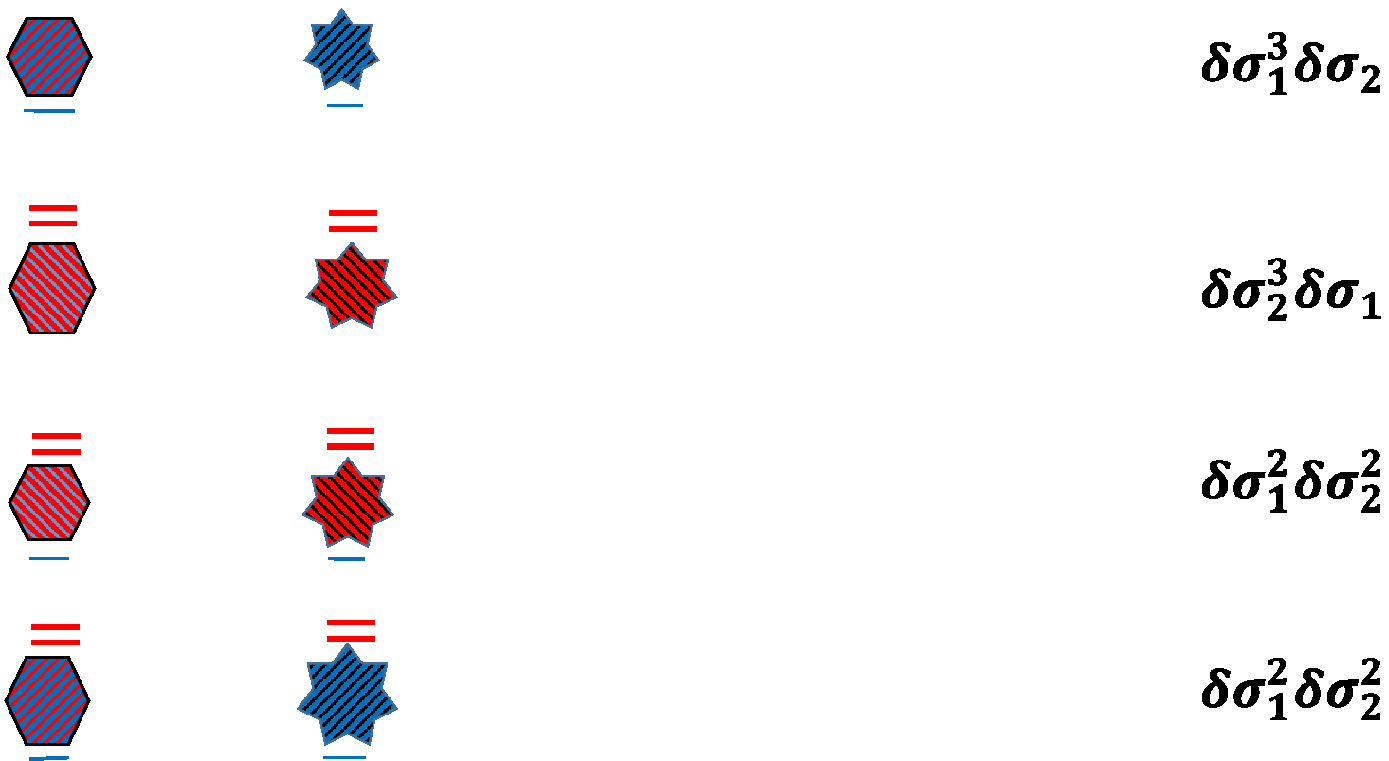}
\label{diagthreepointmain}
\end{figure}

\newpage
\section{Introduction}
\label{sec_intro}

Inflation theory is one of the pioneering models in describing the initial evolutions of the Universe. This model is able to shed light on how large cosmic structures are formed. It is also in accordance with cosmological observations. We should draw attention to this point that the dynamics of inflation theory and how it starts still remains a mystery \cite{Kazanas:1980tx,Guth:1980zm,Starobinsky:1980te,Sato:1980yn,Starobinsky:1982ee,
Linde:1981mu}.
After more than three decades of extensive investigations, the inflationary paradigm is
considered to be a corner stone of the standard model of cosmology, in addition to solving the flatness,  the horizon and the relic problems
\cite{Guth:1980zm,Starobinsky:1980te,Starobinsky:1982ee,
Linde:1981mu}.
In addition, inflation is needed in order to predict the correct behavior of primordial
fluctuations and a Universe albeit with an almost nearly scale-invariant density power
spectrum
\cite{Liddle:1999mq,Langlois:2004de,Lyth:1998xn,Guth:2000ka,Lidsey:1995np,Bassett:2005xm},
 as well as with the correct quantity of tensor perturbations
\cite{Grishchuk:1974ny,Starobinsky:1979ty,Allen:1987bk,Sahni:1990tx,Souradeep:1992sm,
Giovannini:1999qj,Sami:2004xk,Hossain:2014zma,Geng:2015fla,Hossain:2014coa,Cai:2014uka}.
One of the most important duties for cosmologists is an introduction of various models for inflation, then examining the validity of the predictions made and adapting them with observational data. In fact there are two main  methods that one able to  obtain the fulfillment of the
inflationary prototype. In one side, one can consider a modification of the geometrical
sector  leads to a modified general relativity
behavior that allows for inflationary solutions \cite{Nojiri:2003ft,Nojiri:2005pu,Capozziello:2005tf,Capozziello:2011et,Sheikhahmadi:2016wyz}. The most well known proposal in this
method is the Starobinsky inflation \cite{Starobinsky:1980te}. In the other procedure one can
introduce new forms of matter which is able to drive inflation as well. In this approach one usually considers a canonical scalar field,
assuming it to take large values  \cite{Linde:1983gd}
or small values
\cite{Albrecht:1982wi,Freese:1990rb}, a phantom field
\cite{Piao:2004tq,Lidsey:2004xd,Elizalde:2008yf,Feng:2010ya}, a tachyon field
\cite{Fairbairn:2002yp,Feinstein:2002aj,Aghamohammadi:2014aca}, or other
models including k-inflation \cite{ArmendarizPicon:1999rj,Sheikhahmadi:2015gaa} and ghost
inflation \cite{ArkaniHamed:2003uz} etc.
 One can put the latter type of above models in a general set namely single-field inflation model. The physics of single-field model is well investigated \cite{Chen:2009we,Chen:2009zp,Baumann:2011nk,Chen:2012ge, Sefusatti:2012ye, Norena:2012yi, Noumi:2012vr, Gong:2013sma, Emami:2013lma, Kehagias:2015jha, Arkani-Hamed:2015bza,  Dimastrogiovanni:2015pla,Chen:2015lza, Chen:2016cbe,Lee:2016vti,Chen:2016qce, Meerburg:2016zdz,Chen:2016uwp,Chen:2016hrz,
An:2017hlx,An:2017rwo,
Iyer:2017qzw, Kumar:2017ecc, Franciolini:2017ktv, Tong:2018tqf,MoradinezhadDizgah:2018ssw,Saito:2018xge,Chen:2016nrs,Chen:2011zf,Chen:2011tu,Chen:2012ja,Chen:2014joa,Chen:2014cwa,Chen:2017ryl,Chen:2018sce}; however,  some results about inflationary predictions, originated from observational data, might not be accounted for cosmic evolutions regarding only single-field models,  therefore based on the following results multifield models  are afforded with priority. In other word, multifield models of inflation are in a good agreement with the results originated from Planck data \cite{Ade:2015lrj,PhysRevLett112011302}. In Multifield models for instance the inflation can be driven by two fields  instead of one field. In this scenario to investigate quantum  fluctuations and the effects of anisotropy of  temperature observed in CMB only a linear combination was sufficient. But, examining the production of  isocurvature perturbations  and other cases in this veins asks for non-linear combination. In this mode, you can divide the field space into two directions, including inflationary direction and isocurvature one. Similar to what that has done in related literature, we are about to investigate a model in which one direction is related to light field so-called inflaton and other directions related to heavy masses within masses around Hubble parameters. We call this model in its resemblance to quasi-single-field, quasi-single-multifields(QSMF) inflation. one can find relevant calculations due to quasi-single-field model in more details in  references  \cite{Chen:2009we,Chen:2009zp,Baumann:2011nk,Chen:2012ge, Sefusatti:2012ye, Norena:2012yi, Noumi:2012vr, Gong:2013sma, Emami:2013lma, Kehagias:2015jha, Arkani-Hamed:2015bza,  Dimastrogiovanni:2015pla,Chen:2015lza, Chen:2016cbe,Lee:2016vti,Chen:2016qce, Meerburg:2016zdz,Chen:2016uwp,Chen:2016hrz,
An:2017hlx,An:2017rwo,Chen:2017ryl,Chen:2018sce}.
Among the reasons that we can justify the existence of multiple model, we can refer to high energy (specifically on the ground of string theory) as well as non-Gaussianity\cite{Bartolo:2004if,Liguori:2010hx,Chen:2010xka,Wang:2013eqj,Chen:2009we}. In fact, super gravity and string theory related to inflation, we can realize that the field acquires the massive background within Hubble's parameters.
It is therefore justifiable if one or several semi heavy fields similar to that  light field so as to make inflation more plausible. On the other hand, non-Gaussianity is one of the predictions has made by inflation theory. Investigating the size and shape of this quantity is to a great extend can help us in predicting the models which generalize its ability to observe and is the major task of inflation theory. Investigating bispectrum, especially trispectrum based on in-in formulation is conducted comprehensively in recent works and articles about single-field models. Furthermore, following this model,  author in \cite{Emami:2013lma} has conducted multifield that contains abstruse mathematics. Besides this conventional calculating methods, reference author's \cite{Chen:2017ryl} used diagrammtic method and some methods based on Schwinger-Keldysh (S-K) formalism which greatly reduces the amount of computation and gives the same results for many different cases. In this veins, we want to develop the mentioned items based on this new formulation and divide it into multifields \cite{Emami:2013lma} and make a comparison with the aforementioned results. As can seen from this type of interaction related to this action, in conventional computing especially trispectrum is way beyond and sometimes it will be divergent for asymptotic states, therefore, the diagrammatic method could be efficient and economical for this model. So, with development of the aforementioned models (perhaps), introducing quantitatively some new concepts, specifically drawing new diagrams, we aim at investigating the size and shape of a non-Gaussianity in a QSMF model.\\
This paper is organized as follows: In Sec.\,\ref{DiaandMixPro}, we will reintroduce the main  rules  based on S-K formalism and also the diagrammatic and propagators will be expressed as well. In Sec.\,\ref{sec_QSFI} we shall try to explain how the diagrammatic method, for QSMF, leads to tidy results for three-point and four-point correlation functions of primordial scalar perturbations. Sec.\,\ref{BISPECTRUM--} can be considered as a good example to show the increasing of amount of non-Gaussianity in QSMF inflation comparing Quasi single one. For more investigation and also to see the power of the diagrammatic method Sec.\,\ref{Trispectrum} will express the trispectra for QSMF model. At last, Sec.\,\ref{seccon} is devoted to conclusion and discussions.

\section{Diagrammatic Rules and Mixed Propagator}
\label{DiaandMixPro}
In this section following the method introduced in \cite{Chen:2017ryl,Chen:2018sce} for quasi-single-field inflation, but, regarding the QSMF inflationary regime we want to review briefly the S-K formalism and propagators.  The diagrammatic approach based on in-in formalism is reviewed in \cite{Chen:2017ryl,Chen:2018sce}  as well. Additionally, it can be showed that  the diagrammatic method presented in \cite{Chen:2017ryl,Chen:2018sce} is very close to the usual Feynman rules in diagrammatic method, except that the space and the time are treated the same, and that here the model contains two types of propagators introduced by either $+$ or $-$. Accordingly, as it was mentioned in \cite{Chen:2018sce} we know the internal vertex in the diagrams is associated to conserved three-momentum and also an integration of time, and to consider the effects of all interactions appearing in the Lagrangian one should sum over diagrams for all vertices with all possible types of them.
To do this end, one has to  transform the spatial directions but not the temporal one, by virtue of Fourier transformation. Hence, the propagator for a general scalar field $\varphi$ would depends on the time variables of both ends, together with the three-momentum it carries. To receive the propagator, we can use the Fourier-decomposing to the scalar field $\varphi$ based on the related mode function $u(\tau,k)$, viz.
\bge
\label{phimode}
  \varphi (\tau,\mb x)=\int\FR{\di^3\mb k}{(2\pi)^3}\Big[u (\tau, k)b (\mb k)+u^*(\tau, k)b^\dag(-\mb k)\Big]e^{\ii\mb k\cdot\mb x},
\ede
where $\tau$ is the conformal time, $b$ and $b^{\dag}$ are annihilation and creation operators respectively. Thence the evolution equation can be rewritten in terms of $u(\tau,k)$. Accordingly, for the mode function $u(\tau,k)$ with mass $m$ one has
 \begin{equation}\label{eom}
u''(\tau,k)-\frac{2}{\tau}u'(\tau,k)+\bigg(k^2+\frac{m^2}{a(\tau)^2}\bigg)u(\tau,k)=0,
\end{equation}
in which the scale factor is introduced as
$$a(\tau)\simeq -1/(H\tau).$$
Considering the early Universe conditions, the solution of the above equation can be worked out as
\begin{align}
\label{modems}
  u (\tau,k)=-\FR{\ii\sqrt{\pi}}{2}e^{\ii\pi(\nu /2+1/4)}  H(-\tau)^{3/2}\text{H}_{\nu}^{(1)}(- k\tau),
\end{align}
where $\text{H}_\nu^{(1)}(z)$ is the Hankel function of the first kind. When the mass tends to zero, $m=0$, the mode function could be obtained as
\bge
\label{modeml}
    u (\tau,k)=\FR{H}{\sqrt{2k^3}}(1+\ii  k \tau)e^{-\ii  k\tau}.
\ede
To determine the definitions of the tree level propagators  we can consider  the following two-point functions, $\Gamma_{pn}$, \cite{Chen:2017ryl,Chen:2018sce} , namely
\begin{align}
-\ii \Gamma_{pn}(\tau_1,\mb x_1;\tau_2,\mb x_2)=\FR{\de}{\ii p \de J_{p}(\tau_1,\mb x_1)}\FR{\de}{\ii n \de J_{n}(\tau_2,\mb x_2)}Z_0[J_p,J_n]\bigg|_{J_{pn}=0},
\end{align}
where $pn$ refers to $\pm$. Now by regarding different choices for $p, n$ indices, obviously there are four types of propagators. For example, the pp-type propagator can be worked out as,
\label{PropX}
\begin{align}
-\ii \Gamma_{pp}(\tau_1,\mb x_1;\tau_2,\mb x_2)
=&~\FR{\de}{\ii \de J_p(\tau_1,\mb x_1)}\FR{\de}{\ii \de J_p(\tau_2,\mb x_2)}Z_0[J_p,J_n]\bigg|_{J_{pn}=0}\n\\
=&~\int\mathcal{D}\varphi_p\mathcal{D}\varphi_n\,\varphi_p(\tau_1,\mb x_1)\varphi_p(\tau_2,\mb x_2)e^{\ii\int\di\tau\di^3\mb x\,(\ld_0[\varphi_p]-\ld_0[\varphi_n])}\n\\
=&\sum_{\al}\la\Omega|O_\al\ra\la O_\al|\text{T}\{\varphi(\tau_1,\mb x_1)\varphi(\tau_2,\mb x_2)\}|\Omega\ra\n\\
=&~\la\Omega|\text{T}\{\varphi(\tau_1,\mb x_1)\varphi(\tau_2,\mb x_2)\}|\Omega\ra.
\end{align}
Now, to solve above equation and also to clarifying the concept of interaction in the both Quasi-single-field and QSMF  let us introducing two types of Lagrangian as follows
\bge
\ld_\text{cl}[\varphi]=\ld_0[\varphi]+\ld_\text{int}[\varphi],
\ede
 here $\ld_0$ refers free part and $\ld_\text{int}$ indicates the interaction terms. According to above discussions, the term $\ld_0$  contains all quadratic terms  in terms of  $\varphi$ and all remaining terms could be put in $\ld_\text{int}$ one.  As it will be seen, this latter contains a rich physics and leads to very important predictions about non-Gaussianity as well.  To complete this brief review, considering concepts risen from (effective) quantum field theory and by following our main reference  \cite{Chen:2018sce} we can define the generating functional $Z[J_p J_n]$ as
\begin{align}
\label{GF}
 Z[J_p J_n]=\int\mathcal{D}\varphi_{p} \mathcal{D}\varphi_{n} \,
\exp\bigg[\ii\int_{\tau_0}^{\tau_f}\di\tau\di^3\mb x\,\Big(\ld_\text{cl}[\varphi_p]-\ld_\text{cl}[\varphi_n]+J_p\varphi_p-J_n\varphi_n\Big)\bigg].
\end{align}
Beside, by taking functional derivative and considering general amplitude
$$\la \varphi_{a_1}(\tau,\mb x_1)\cdots\varphi_{a_N}(\tau,\mb x_N)\ra(a_1,\cdots,a_N =\pm)$$
at last the four mentioned propagators could be obtained as follows
\bgs
\label{SKprop}
\begin{align}
G_{pp}(k;\tau_1,\tau_2)=&~G(k;\tau_1,\tau_2)\Theta(\tau_1-\tau_2)+\bar{G}(k;\tau_1,\tau_2)\Theta(\tau_2-\tau_1),\\
G_{pn}(k;\tau_1,\tau_2)=&~\bar{G}(k;\tau_1,\tau_2),\\
G_{np}(k;\tau_1,\tau_2)=&~G(k;\tau_1,\tau_2),\\
G_{nn}(k;\tau_1,\tau_2)=&~\bar{G}(k;\tau_1,\tau_2)\Theta(\tau_1-\tau_2)+G(k;\tau_1,\tau_2)\Theta(\tau_2-\tau_1),
\end{align}
\eds
here
\bgs
\label{Ggs}
\begin{align}
G(k;\tau_1,\tau_2)\equiv&~u(\tau_1,k)u^*(\tau_2,k),\\
\bar{G}(k;\tau_1,\tau_2)\equiv&~u^{*}(\tau_1,k)u(\tau_2,k).
\end{align}
\eds
where $\Theta(z)$ is the step function. To calculate the required three-point function of $\de\phi$ with intermediate exchange of $\de\si$ one would regard the propagators (\ref{SKprop}) and the mode functions (\ref{modems}) or (\ref{modeml}), and follow the diagrammatic rules as reviewed in \cite{Chen:2017ryl,Chen:2018sce}. Based on diagrammatic rules were presented in Chen et al. \cite{Chen:2017ryl}, the three-momentum related to the self interacting  cases could be summarized as bellow,
\begin{figure}[ht]
\centering
\includegraphics[scale=.950]{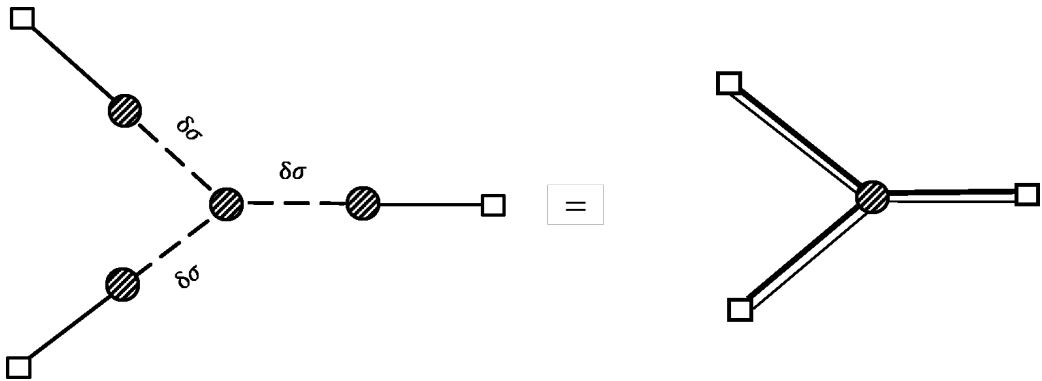}
\caption{{\it{In this diagram the solid lines express the propagators of massless field $\de\phi$ while dashed lines represent the propagator of massive fields $\de\si_i$, and $i$ refers the different components of these fields.}}}
\label{diagthreepointmain}
\end{figure}
  Also from figure \ref{diagthreepointmain} it is realized that the external points are distinguished by little squares, while the shaded dots in each vertex means that it is related to two different types of propagators. The black circles refer to the plus propagators  and the minus propagators indicated by white circles, and one should sum over all possible states.  Accordingly, figure \ref{diagthreepointmain} actually represents the sum of 16 different diagrams since we have 4 shaded circles. Albeit one should keep in mind that each black circle is complex conjugate of white one and vice versa. This fact dramatically decreases the amount of calculations, the calculation can be simplified by noting to the repetitive structure on the left-hand side of figure \ref{diagthreepointmain}, this approach completely explained in \cite{Chen:2017ryl,Chen:2018sce}.

\section{Application to QSMF Inflation}
\label{sec_QSFI}

In this section, we consider QSMF Inflation \cite{Emami:2013lma}, by following quasi-single-field model \cite{Chen:2009we,Chen:2009zp} and \cite{Chen:2017ryl,Chen:2018sce}, as an example to show how the diagrammatic method leads to relatively compact results compared to standard in-in formalism for three-point and four-point correlation functions of primordial scalar perturbations. As explained in aforementioned sections, the  QSMF Inflation in general refers to the inflation scenarios with two or more spectator fields of mass around Hubble scale which are coupled kinetically to the inflaton, light field beside the internal interactions. To do so, we consider the following action with a slightly curved inflation trajectory described by  three real scalar fields $\theta$, $\si_{1}$ and $\si_{2}$,
\begin{eqnarray}
  S&=& \int\di^4x\sqrt{-g}\times\\ \nonumber
  & &\bigg[-\FR{1}{2}(\wt{R}+ \sigma_1 + \sigma_2)^2(\pd_\mu\theta)^2- \FR{1}{2}(\pd_\mu\si_1)^2-\FR{1}{2}(\pd_\mu\si_2)^2
   -V_{\text{sr}}(\theta)-V(c_1 \sigma_1 + c_2 \sigma_2)\bigg]\, ,
\label{Action-a}
\end{eqnarray}
where $V_\text{sr}(\theta)$ is an arbitrary slow-roll potential, while $V(\si)$ is a potential such that $\si$ obtains a classical constant background $\si_0$. After expanding the fields around their classical background $\theta_0$ and $\si_0$, the Lagrangian for the fluctuation field has the following form,
\begin{align}\label{QBFL}
\ld_\text{cl}=\ld_0+\ld_1+\ld_2,
\end{align}
 where
\[\ld_0=\Big[\FR{a^2}{2}\Big((\de\phi')^2-(\pd_i\de\phi)^2+(\de\si_1')^2(\de\si_2')^2-(\pd_i\de\si_1)^2-(\pd_i\de\si_2)^2\Big)\n\\
-\FR{a^4m_1^2}{2}\de\si_1^2-\FR{a^4m_2^2}{2}\de\si_2^2\Big],\]
$$\ld_1=a^3\lam_{21}\de\si_1\de\phi'+a^3\lam_{22}\de\si_2\de\phi'
-a^4\Big(\FR{\lam_{31}}{6}\de\si_1^3+\FR{\lam_{32}}{6}\de\si_2^3+\FR{{\Lambda}_{31}}{6}\de\si_1^2\de\si_2+\FR{{\Lambda}_{32}}{6}\de\si_2^2\de\si_1+\Big),$$
and
\begin{eqnarray}\n
\ld_2&=&-a^4\Big(\FR{\lambda_{41}}{24}\de\si_1^4+\FR{\lambda_{42}}{24}\de\si_2^4+\FR{{\Lambda}_{41}}{24}\de\si_1^3\de\si_2+\FR{{\Lambda}_{42}}{24}\de\si_2^3\de\si_1
+\FR{\tilde{\Lambda}_{4j}}{24}\de\si_1^2\de\si_2^2+\cdots\Big)\\\n
&+&\frac{a^3\dot{\phi_0}}{R^2}\big(\de\si_1^2\de\phi'+\de\si_2^2\de\phi'\big)+a^2\big(\frac{\de\si_j}{R}+\frac{\de\si_j^2}{2R^2}\big)\big[(\de\phi')^2-(\pd_i\de\si_2)^2\big].
\end{eqnarray}
Where we have defined $\de\phi=(\wt R+\si_{01}+\si_{02})\de\theta$. The first term of the above Lagrangian can be justified as free part $\ld_0$, with a massless scalar $\de\phi$ and two massive scalar fields $\de\si$ of mass functions $m_1^2=V''(\si_{01})-\dot\theta_0^2$, and $m_2^2=V''(\si_{02})-\dot\theta_0^2$. In $\ld_1$, we have interactions with two-point derivative mixing between $\de\phi$ and $\de\si$, with coupling strength $\lam_{2j}=2\dot\theta_0$, as well as self-interactions of $\de\si$, with couplings $\lam_{3j}=V'''(\si_{0j})$, ${\Lambda}_{31}\equiv3\frac{\partial}{\partial \si_2}(\frac{\partial^2}{\partial^2 \si_1}V)$ and ${\Lambda}_{32}\equiv3\frac{\partial}{\partial \si_1}(\frac{\partial^2}{\partial^2 \si_2}V)$. In a same procedure, for $\ld_2$ we can write
 $\lam_{4j}=V^{(4)}(\si_{0j})$, ${\Lambda}_{41}\equiv4\frac{\partial}{\partial \si_2}(\frac{\partial^3}{\partial^3 \si_1}V)$, ${\Lambda}_{42}\equiv4\frac{\partial}{\partial \si_1}(\frac{\partial^3}{\partial^3 \si_2}V)$ and $\tilde{\Lambda}_{4j}\equiv6(\frac{\partial^2}{\partial^2 \si_1}(\frac{\partial^2}{\partial^2 \si_2}V)$,
 In above equations $V^{(4)}=V''''$.
Our main mean to extend the $\ld_\text{cl}$ up to $4th$ order is to see the effects of extra massive fields on the behaviour of bispectrum and especially trispectrum in QSMF proposal. Additionally, we should exceed the primary orders in leading terms to see obviously the  effects of S-K diagrammatic approach  on decreasing the amount of calculations. But at first let us warm up by computing the power spectrum and then the corrections, due to the extra massive fields, will be eye-catching. After that, we shall calculate both  bispectram and trispectrum as well, for more detail one can see \cite{Chen:2017ryl,Chen:2018sce} .
 Accordingly, before any computation of various correlation functions we want to study the evaluation of a mixed propagator for at hand QSMF model in more details next subsection.

\subsection{Mixed Propagator in QSMF model}

Whereas a special type of interacting terms appears in different places in our investigation so following  \cite{Chen:2017ryl,Chen:2018sce}  we are going to introduce this special case as mixed propagator. In fact for the QSMF inflation model we shall calculate the contribution of interaction of both massive fields with inflaton and therefore the following objects can be drawn

\begin{figure}[ht]
\centering
\includegraphics[scale=.80]{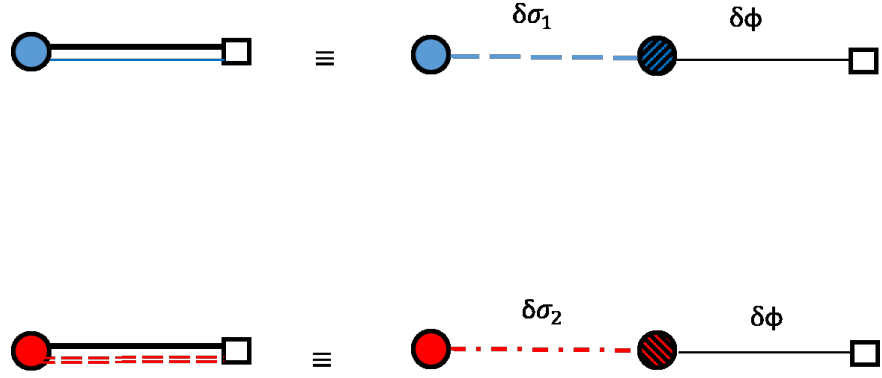}
\caption{{\it{The first diagram expresses  an equivalent for two point function  $\la\de\si_1\de\phi\ra$. the dashed (blue) lines express the propagators of massive field $\de\si_1$. The latter diagram presents  an equivalent for two point function  $\la\de\si_2\de\phi\ra$. the dotted dashed (red) lines express the propagators of massive field $\de\si_2$.}}}
\label{Sig1}
\end{figure}

\begin{figure}[ht]
\centering
\includegraphics[scale=.80]{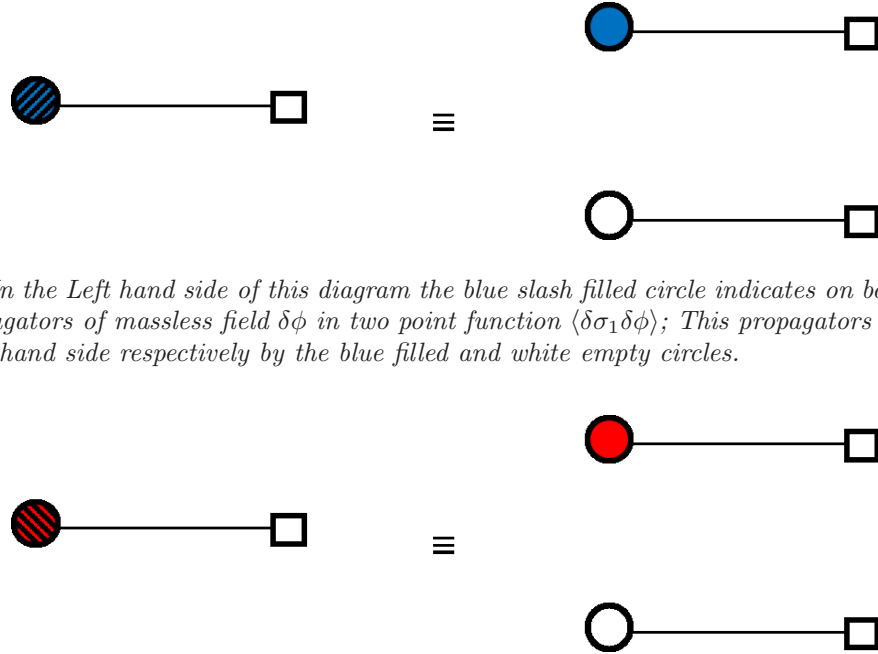}
\caption{{\it{In the Left hand side of this diagram the blue slash filled circle indicates on  both types of $+$ and $-$ propagators of massless field $\de\phi$ in two point function  $\la\de\si_1\de\phi\ra$; This propagators are presented in the right hand side respectively by the blue filled and white empty circles.}}}
\label{Sig1-2}
\end{figure}

\begin{figure}[ht]
\centering
\includegraphics[scale=.80]{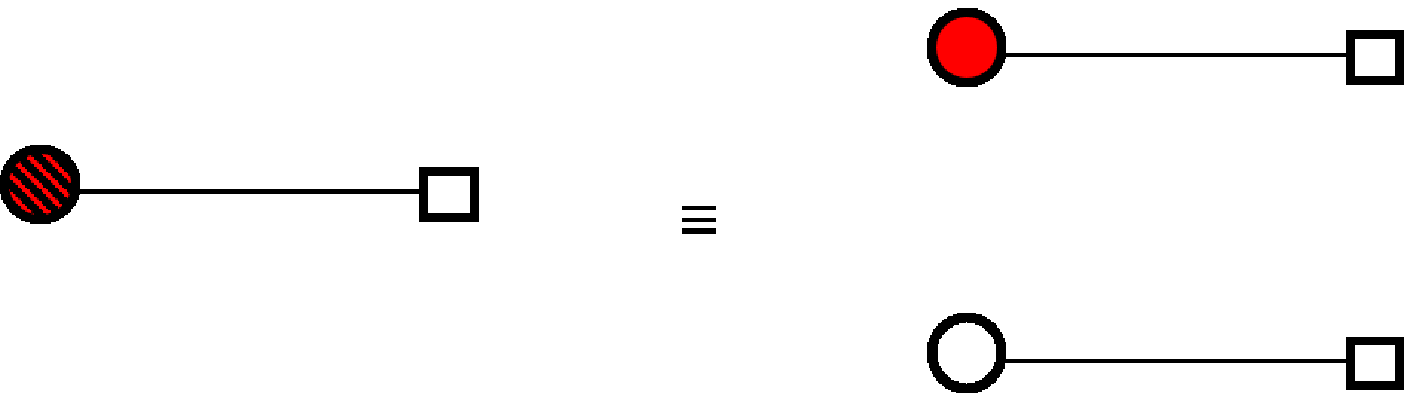}
\caption{{\it{In the Left hand side of this diagram the red backslash filled circle indicates on  both types of $+$ and $-$ propagators of massless field $\de\phi$ for two point function  $\la\de\si_2\de\phi\ra$; They are presented in the right hand side respectively by the red filled and white empty circles.}}}
\label{Sig1-21}
\end{figure}
\begin{figure}[ht]
\centering
\includegraphics[scale=.80]{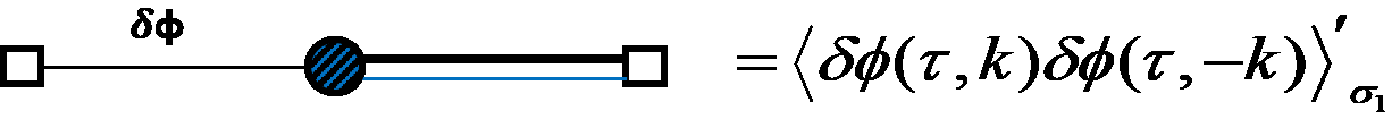}
\caption{{\it{In this diagram we present the leading order correction to the power spectrum from two-point function based on Eq.(\ref{2-point}) for the massive field $\delta\sigma_1$.}}}
\label{sigma1leading}
\end{figure}
\begin{figure}[ht]
\centering
\includegraphics[scale=.80]{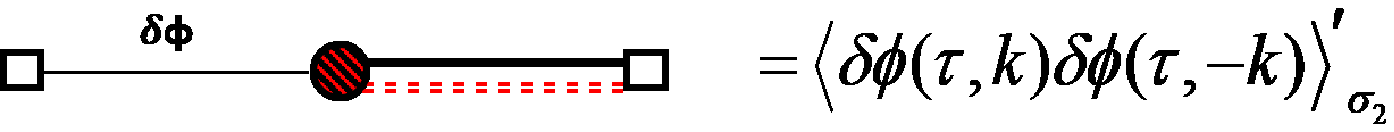}
\caption{{\it{In this diagram we show the leading order correction to the power spectrum from two-point function based on Eq.(\ref{2-point}) for the massive field $\delta\sigma_2$.}}}
\label{sigma2leading}
\end{figure}

Figures \ref{Sig1}- \ref{Sig1-21} could be considered as an equivalent for two point function  $\la\de\si_j\de\phi\ra$, in which $\de\phi$ refers the  values of scalar field perturbations in present era. Now for massive and light fields we should use their own propagators and therefore for example the  propagator of $\de\si_j$ can be denoted by $D_j(k;\tau_1,\tau_2)$  and the propagator of  $\de\phi$ by $G(k;\tau_1,\tau_2)$ see figures \ref{sigma1leading} and \ref{sigma2leading}. Accordingly
\begin{align}\label{2-point}
\mathcal{G}_{\pm j}(k;\tau)=&~\ii\lam_{2j} \int_{-\infty}^0 \FR{\di\tau'}{(-H\tau')^3}\Big[D_{j\pm+}(k;\tau,\tau')\pd_{\tau'}G_{+}(k;\tau')-D_{j\pm-}(k;\tau,\tau')\pd_{\tau'}G_{-}(k;\tau')\Big]\n\\
=&~\sum_{j=1}^{2}\FR{\pi\lam_{2j} H}{8k^3}I_{\pm j}(z),
\end{align}
where $z\equiv -k\tau$, and $I_\pm(z)$ are expressed by,
\begin{align}
\label{Ipm}
I_{\pm j}(z)=&~e^{-\pi\,\text{Im}\,{{\nu_j}}}z^{3/2}\bigg\{2\,\text{Im}\bigg[\mathrm{H}_{{\nu_j}}^{(1)}(z)\int_0^{\infty}\FR{\di z'}{\sqrt{z'}} \mathrm{H}_{{{{\nu_j}}}^*}^{(2)}(z')e^{-(\ii +\ep) z'}\bigg]\n\\
&~+\ii \mathrm{H}_{{\nu_j}}^{(1)}(z)\int_0^z\FR{\di z'}{\sqrt{z'}} \mathrm{H}_{{{\nu_j}}^*}^{(2)}(z')e^{\mp\ii  z'}-\ii \mathrm{H}_{{{\nu_j}}^*}^{(2)}(z)\int_0^z\FR{\di z'}{\sqrt{z'}} \mathrm{H}_{{\nu_j}}^{(1)}(z')e^{\mp\ii  z'}\bigg\},
\end{align}
where $\ep$ is a small constant. After some algebra for the above integral one obtains
\begin{align}
I_{\pm j}(z)
=&~z^{3/2}e^{-\pi\,\text{Im}\,{{\nu_j}}}\bigg\{\Big[\mathcal{C}_{{\nu_j}}+(\cot(\pi{{\nu_j}})-\ii)f_{{\nu_j}}^\pm(z)-\csc(\pi{{\nu_j}})f^\pm_{-{{\nu_j}}}(z)\Big]\mathrm{H}_{{{\nu_j}}^*}^{(2)}(z)\n\\
&~+\Big[\mathcal{C}_{{\nu_j}}^*+(\cot(\pi{{\nu_j}}^*)+\ii)f^\pm_{{{\nu_j}}^*}(z)-\csc(\pi{{\nu_j}}^*)f^\pm_{-{{\nu_j}}^*}(z)\Big]\mathrm{H}_{{{\nu_j}}}^{(1)}(z)
\bigg\},
\end{align}
where $f^\pm_{{\nu_j}}(z)$ is defined by,
\begin{align}
\label{fnu}
  f^\pm_{{\nu_j}} (z)=\FR{z^{{{\nu_j}}+1/2}}{2^{{\nu_j}}({{\nu_j}}+1/2)\Gamma({{\nu_j}}+1)}{}_2F_2\Big({{\nu_j}}+\FR{1}{2},{{\nu_j}}+\FR{1}{2};{{\nu_j}}+\FR{3}{2},2{{\nu_j}}+1;\mp 2\ii z\Big).
\end{align}
In the $z\rightarrow +\infty$ limit of hypergeometric function, by virtue of the asymptotic behavior
\bea
&& {}_2F_2(a,a;b_1,b_2;z) \xrightarrow{|z|\to \infty} \frac{\Gamma(b_1)\Gamma(b_2)}{\Gamma(a)^2} e^z z^{2a-b_1-b_2}
\nonumber \\
&& + \frac{\Gamma(b_1) \Gamma(b_2)}{\Gamma(a) \Gamma(b_1-a) \Gamma(b_2-a)}
(-z)^{-a}
\left( \ln(-z) - \psi(b_1-a) - \psi(b_2-a) - \psi(a) - 2\gamma \right) ~,
\eea
where
\bea
\psi(z) \equiv \frac{d\ln\Gamma(z)}{dz} ~,
\eea
and $\mathcal{C}_{{\nu_j}}$ is a $z$-independent coefficient, given by
\begin{align}
\label{cnu}
\mathcal{C}_{{\nu_j}}=&~\ii\int_0^\infty\FR{\di z}{\sqrt{z}}H_{{\nu_j}}^{(1)}(z)e^{ (\ii +\ep) z}=\sqrt{2\pi}e^{\ii\pi(1/4-{{\nu_j}}/2)}\sec(\pi{{\nu_j}}),
\end{align}
in which  for $\ep\to 0$ limit the UV convergence will be resulted.

Using the diagrammatic rules, we can write down the corresponding expression immediately,
\begin{align}
&\la\de\phi(\tau,\mb k)\de\phi(\tau,-\mb k)\ra'\n\\
=&~\ii\lam_{2 j} \int_{-\infty}^0\FR{\di\tau'}{(-H\tau')^3}\Big[\pd_{\tau'}G_{+}(k;\tau')\mathcal{G}_+(k;\tau')-\pd_{\tau'}G_{-}(k;\tau')\mathcal{G}_-(k;\tau')\Big]\n\\
=&~\FR{\lam_{2 j} ^2}{k^3}\sum_{j=1}^{2}\mathcal{P}({{\nu_j}}),
\end{align}
where $\mathcal{P}({{\nu_j}})$ is the following integral and can be carried out completely as was done in \cite{Chen:2012ge},
\begin{align}\label{Powespectra}
  \sum_{j=1}^{2}\mathcal{P}({{\nu_j}})
\equiv \sum_{j=1}^{2}\FR{-\ii \pi}{16}\int_0^\infty\FR{\di z}{z^2}\Big[e^{-\ii z}I_{+ j}(z)-e^{+\ii z}I_{- j }(z)\Big]\n\\
  = \sum_{j=1}^{2}\FR{\pi^2}{4\cos^2(\pi{{\nu_j}})}+\Xi({{\nu_j}})+\Xi(-{{\nu_j}}),
\end{align}
and here the function $\Xi({{\nu_j}})$ is defined to be,
\bge\label{Xi-}
  \Xi({{\nu_j}})\equiv
  \text{Im} \left\{
  \FR{e^{-\ii\pi{{\nu_j}}}}{16\sin(\pi{{\nu_j}})}\bigg[\psi^{(1)}\Big(\FR{1}{4}+\FR{{{\nu_j}}}{2}\Big)-\psi^{(1)}\Big(\FR{3}{4}+\FR{{{\nu_j}}}{2}\Big)\bigg]\right\},
\ede
 and  $\psi^{(1)}(z)\equiv\di^2\log\Gamma(z)/\di z^2$ is well know PolyGamma function.

 \begin{figure}[ht]
\centering
\includegraphics[scale=.80]{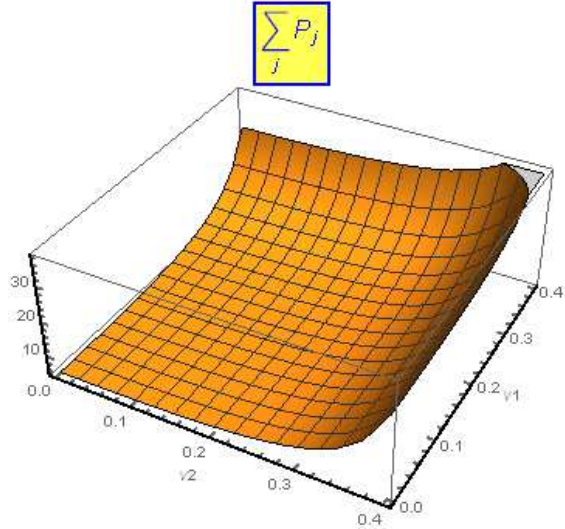}
\caption{{\it{This figure shows the behavior of function of power spectrum  $\mathcal{P}({{\nu_j}})$ introduced in Eq.(\ref{Powespectra}). In this frame we consider same quantities for parameters $\nu_1$ and $\nu_2$.}}}
\label{PSI3}
\end{figure}

 \begin{figure}[ht]
\centering
\includegraphics[scale=.80]{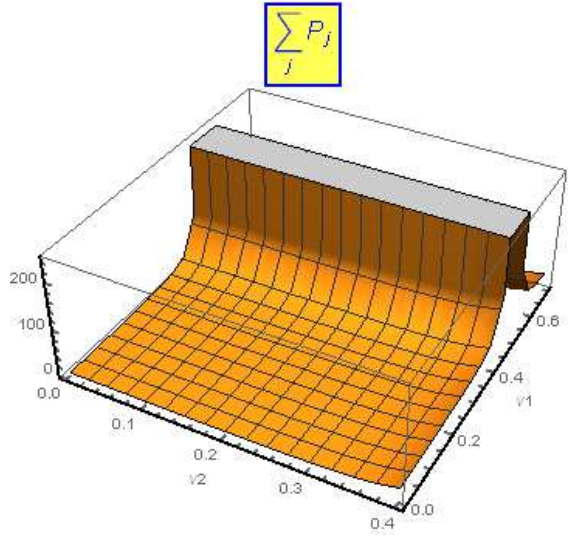}
\caption{{\it{In this diagram we want to examine the effects of  changing the quantity of parameters $\nu_j$ on the behaviour of power spectrum. We saw that if  one consider different ranges for $\nu_1$ and $\nu_2$  it causes to increase the amount of power spectrum.}}}
\label{PSI3Aa}
\end{figure}

 \begin{figure}[ht]
\centering
\includegraphics[scale=.90]{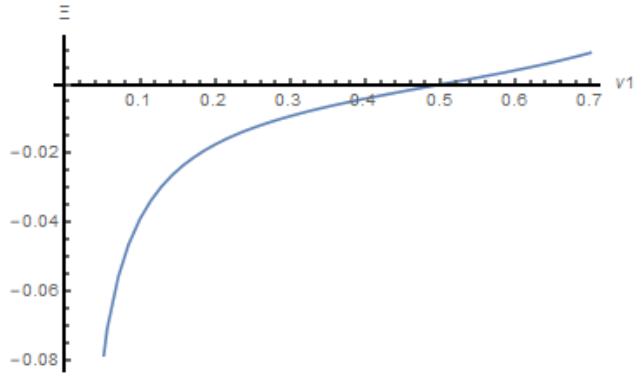}
\caption{{\it{In this diagram the solid line expresses the amount of  $\Xi({{\nu_1}})$ in \ref{Xi-} based on parameter $\nu_1$. It will be realized if $\nu_1\approx0.49$ then   $\Xi({{\nu_1}})$ tends to zero and for greater quantities than $\nu_1\approx0.49$ it enters the positive area.}}}
\label{Theta1}
\end{figure}
From figures \ref{PSI3},  \ref{PSI3Aa} and \ref{Theta1} it is understood that the effects of extra fields in QSMF inflation increase the amount of $\mathcal{P}({{\nu_j}})$ comparing with quasi-single-field model. Albeit we should emphasise that our result has a good agreement, maybe trivially, with the calculations in canonical in-in formalism were done by \cite{Emami:2013lma}.

 \subsection{Bispectrum}
 \label{BISPECTRUM--}
\begin{figure}[ht]
\centering
\includegraphics[scale=.90]{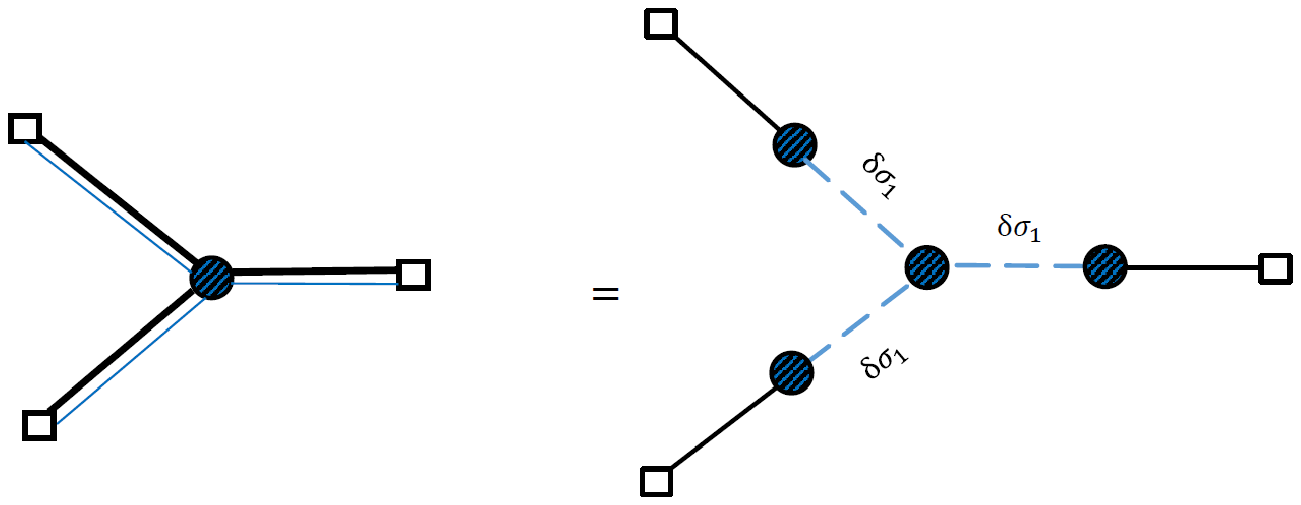}
\caption{{\it{In this diagram based on Eq.(\ref{3pt}) we present the leading orders in expectation value $\la\de\phi(\tau,\mb k_1)\de\phi(\tau,\mb k_2)\de\phi(\tau,\mb k_3)\ra_{\lam_{31}}'$  for $\delta\sigma_1.$}}}
\label{sig11sig22}
\end{figure}

\begin{figure}[ht]
\centering
\includegraphics[scale=.90]{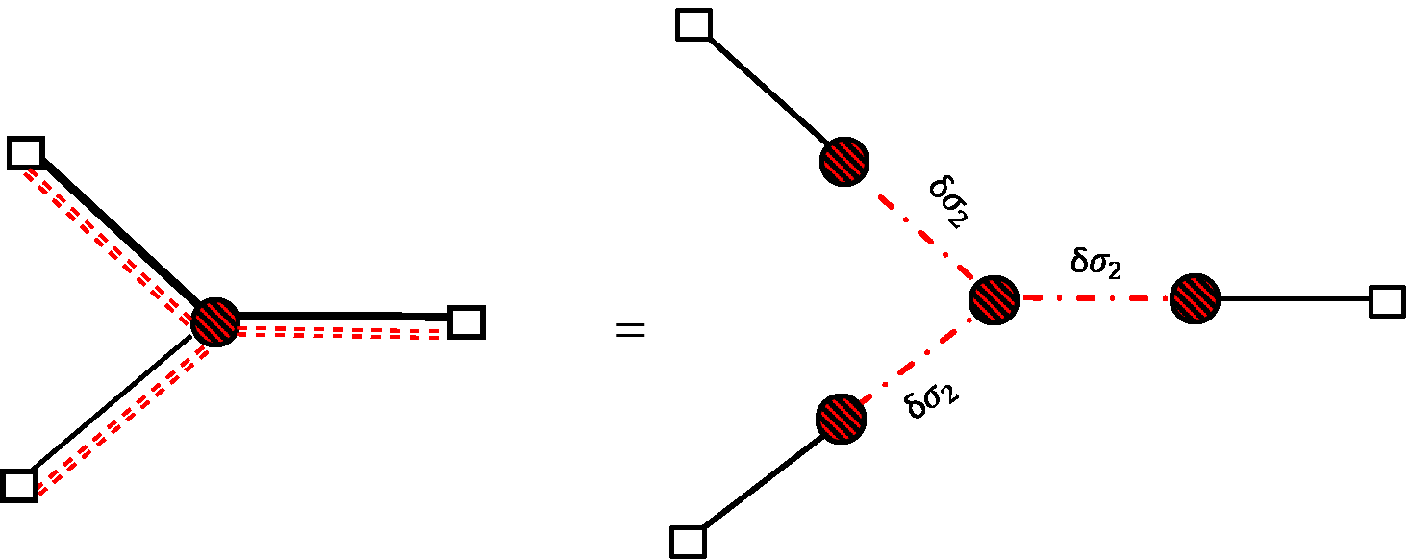}
\caption{{\it{In this diagram based on Eq.(\ref{3pt}) we present the leading orders in expectation value $\la\de\phi(\tau,\mb k_1)\de\phi(\tau,\mb k_2)\de\phi(\tau,\mb k_3)\ra_{\lam_{32}}'$  for $\delta\sigma_2.$}}}
\label{sig11sig22a}
\end{figure}

\begin{figure}[ht]
\centering
\includegraphics[scale=.90]{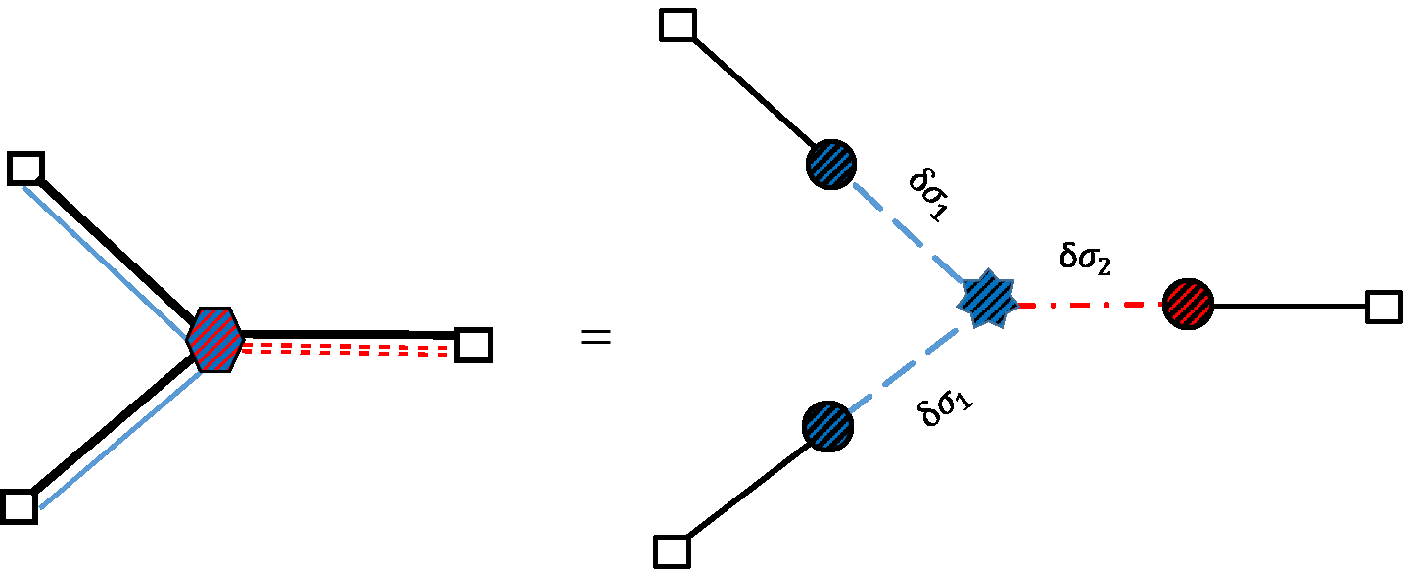}
\caption{{\it{In this diagram based on Eq.(\ref{3pt-b}) we present the leading orders in expectation value $\la\de\phi(\tau,\mb k_1)\de\phi(\tau,\mb k_2)\de\phi(\tau,\mb k_3)\ra_{\Lambda_{3j,l}}'$.}}}
\label{sig12sig21b}
\end{figure}

\begin{figure}[ht]
\centering
\includegraphics[scale=.90]{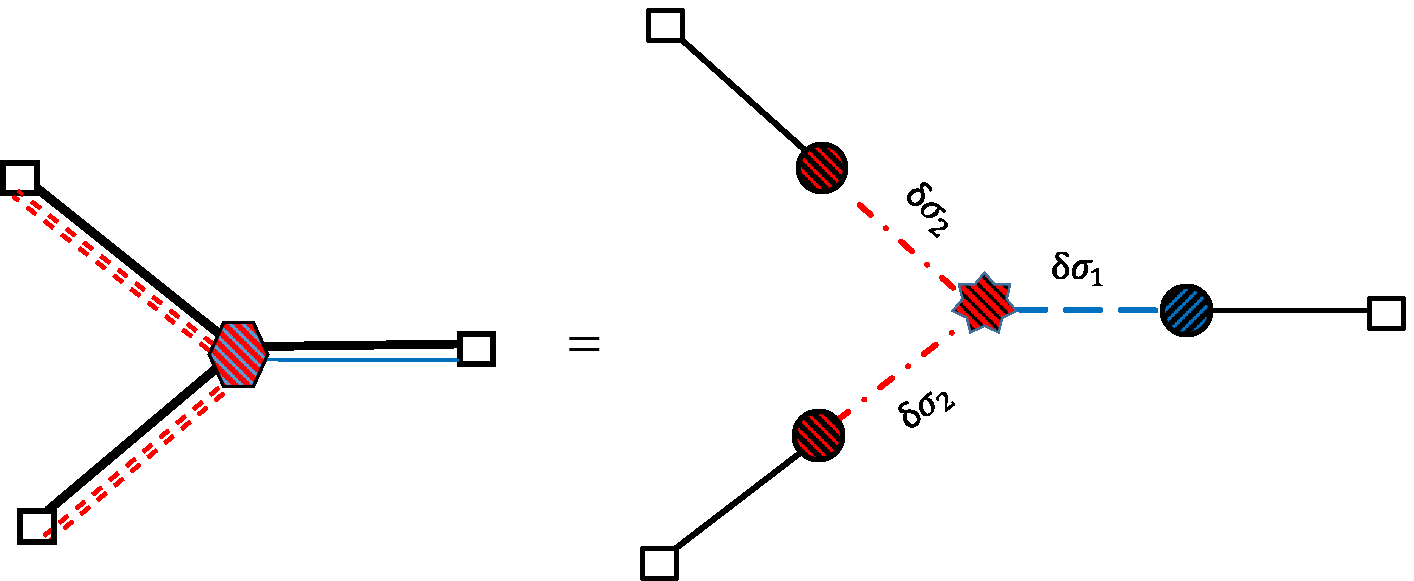}
\caption{{\it{In this diagram based on Eq.(\ref{3pt-b}) we present the leading orders in expectation value $\la\de\phi(\tau,\mb k_1)\de\phi(\tau,\mb k_2)\de\phi(\tau,\mb k_3)\ra_{\Lambda_{3j,l}}'$.}}}
\label{sig12sig21}
\end{figure}

To calculate bispectrum and trispectrum, now we look at the higher order interactions. Following \cite{Chen:2017ryl,Emami:2013lma} and by virtue of the Eqs.(\ref{3pt})  and  (\ref{3pt-b}) and this fact that $R$ can be considered as a cutoff it could be realized that the leading portions to the three-point function are of order $\lam_2^3\lam_3^{}$ and ${\lam_{2j}^2 \lam_{2l} \Lambda_{3j,l}^{}}$. To see how this suitable method actually does work, we can consider the discussion brought in Sec.\ref{DiaandMixPro}. Accordingly, by means of the rules have indicated in \cite{Chen:2016uwp,Chen:2017ryl,Chen:2018sce} and also considering the mixed propagator originating from the Lagrangian  \ref{QBFL} we have the diagrams \ref{sig11sig22}-\ref{sig12sig21}.
From these figures, i.e. \ref{sig11sig22}-\ref{sig12sig21}, it is realized that diagrammatic in-in formalism has more advantages compared to the ordinary in-in formalism. One of this advantages is that the amount of calculations and integrations are dramatically reduced. In fact, each one of the the above diagrams itself contains $16$ diagrams which by virtue of the conjugation rules one only needs to consider these general cases and then sum over all possible states. Another difference between QSMF model and Quasi-single-field, beside self interaction terms,  is the appearance of the interaction terms associated with the massive fields. These extra effects enhances non-Gaussianity so it can be used observationally to distinguish between QSMF and Quasi-single-field scenarios. We have
\begin{align}
\label{3pt}
\la\de\phi(\tau,\mb k_1)\de\phi(\tau,\mb k_2)\de\phi(\tau,\mb k_3)\ra_{\lam_{3j}}'
&=&~\sum_{j=1}^{2}2\lam_{3j}\,\text{Im}\int_{-\infty}^0 \FR{\di\tau}{(-H\tau)^4}\mathcal{G}_{+j}(k_1;\tau)\mathcal{G}_{+j}(k_2;\tau)\mathcal{G}_{+j}(k_3;\tau)\n\\
&=&~\sum_{j=1}^{2}\FR{\pi^3 \lam_{2j}^3 \lam_{3j}^{}}{256Hk_2^3k_3^3}\,\text{Im}\int_0^\infty\FR{\di z}{z^4}I_{+j}(z)I_{+j}(\FR{k_2}{k_1}z)I_{+j}(\FR{k_3}{k_1}z),
\end{align}
and
\begin{align}
\label{3pt-b}
\la\de\phi(\tau,\mb k_1)\de\phi(\tau,\mb k_2)\de\phi(\tau,\mb k_3)\ra_{\Lambda_{3j,l}}'
=&~\sum_{j=1}^{2}\sum_{l=2}^{1}2{\Lambda_{3j,l}}\,\text{Im}\int_{-\infty}^0 \FR{\di\tau}{(-H\tau)^4}\mathcal{G}_{+j}(k_1;\tau)\mathcal{G}_{+j}(k_2;\tau)\mathcal{G}_{+l}(k_3;\tau)\n\\
=&~\sum_{j=1}^{2}\sum_{l=2}^{1}\FR{\pi^3 \lam_{2j}^2 \lam_{2l} \Lambda_{3j,l}^{}}{256Hk_2^3k_3^3}\,\text{Im}\int_0^\infty\FR{\di z}{z^4}I_{+j}(z)I_{+j}(\FR{k_2}{k_1}z)I_{+l}(\FR{k_3}{k_1}z),
\end{align}
where $j, l =1~or~ 2$ and it should be kept in mind when $j=1(2)$ then $l=2(1)$. To calculate the integrals in (\ref{3pt}) and (\ref{3pt-b}) instead of a four-layer integral in the canonical in-in formalism of propagators  one needs to calculate only an one-layer integral \cite{Chen:2017ryl,Chen:2018sce} . To do so one can introduce the dimensionless shape function $S(k_1,k_2,k_3)$ as
\bge
\label{shape}
  \la\zeta(\mb k_1)\zeta(\mb k_2)\zeta(\mb k_3)\ra'\equiv (2\pi)^4S(k_1,k_2,k_3)\FR{1}{(k_1k_2k_3)^2}P_\zeta^2.
\ede
Here $\zeta$ refers the curvature perturbation, which can be expressed as $\zeta =-H\de\phi/\dot\phi_0$, and $P_\zeta=H^2/(8\pi^2\Mp^2\ep)$ is the well-known power spectrum related to  the curvature perturbation.
Following \cite{Chen:2017ryl}, to isolate  the clock signal from the bispectrum we rewrite the expansion of  (\ref{3pt}) and (\ref{3pt-b}) in $k_3/k_1\to 0$ limit and then by virtue of (\ref{shape}) one obtains
\begin{align}
  S_{pure}(k_1,k_2,k_3) \to  P_\zeta^{-1/2}\sum_{j=1}^{2}
  \left(\frac{\lambda_{2j}}{H}\right)^3 \left ( \frac{\lambda_{3j}}{H} \right )\times\n\\
  \text{Im}\,\bigg[
    s_+(\wt{\nu}_{j})\Big(\FR{k_3}{k_1}\Big)^{1/2+\ii\wt{\nu}_{j}}
  + s_-(\wt{\nu}_{j})\Big(\FR{k_3}{k_1}\Big)^{1/2-\ii\wt{\nu}_{j}}\bigg]~~~~~,
\end{align}
and
\begin{align}
 S_{mixed}(k_1,k_2,k_3) \to P_\zeta^{-1/2}\sum_{j=1}^{2}\sum_{l=2}^{1}
  \left(\frac{\lam_{2j}^2 \lam_{2l}}{H^3}\right) \left ( \frac{\Lambda_{3j,l}}{H} \right )\times\n\\
  \text{Im}\,\bigg[
    s_+(\wt{\nu}_ {j,l})\Big(\FR{k_3}{k_1}\Big)^{1/2+\ii\wt{\nu}_ {j,l}}
  + s_-(\wt{\nu}_ {j,l})\Big(\FR{k_3}{k_1}\Big)^{1/2-\ii\wt{\nu}_ {j,l}}\bigg],
\end{align}
where  we would redefine ${{\nu_j}}=\ii\wt{{\nu_j}}$.  It is realized that if one considers $m>3H/2$ limit, then  $\wt{\nu_j}$ could be treated as a real part of equation. Accordingly, for the case at hand  the coefficients $s_\pm(\wt{\nu j})$ are expressed as
\begin{align}
  s_+(\wt{\nu}_j)=&~\sum_{j=1}^{2}\FR{-2^{-\ii\wt{\nu}_ j}\pi^{5/2}}{256\Gamma(1+\ii\wt{\nu}_ j)\sinh(\pi\wt{\nu}_ j)\big[\sinh(\pi\wt{\nu}_ j/2)+\ii\cosh(\pi\wt{\nu_j}/2)\big]}\int_0^\infty\di z\,I_+^2(z)z^{-5/2+\ii\wt{\nu}_ j},\\
  s_-(\wt{\nu}_ j)=&~\sum_{j=1}^{2}\FR{-2^{+\ii\wt{\nu_j}}\pi^{5/2}}{256\Gamma(1-\ii\wt{\nu j})\sinh(\pi\wt{\nu j})\big[\sinh(\pi\wt{\nu}_ j/2)-\ii\cosh(\pi\wt{\nu_j}/2)\big]}\int_0^\infty\di z\,I_+^2(z)z^{-5/2-\ii\wt{\nu_j}},\\
  s_+(\wt{\nu}_ {j,l})=&~\sum_{j=1}^{2}\sum_{l=2}^{1}\FR{-2^{-\ii\wt{\nu_j}}\pi^{5/2}}{256\Gamma(1+\ii\wt{{\nu_j}})\sinh(\pi\wt{\nu}_j)\big[\sinh(\pi\wt{\nu}_ j/2)+\ii\cosh(\pi\wt{\nu}_ j/2)\big]}\int_0^\infty\di z\,I_+^2(z)z^{-5/2+\ii\wt{\nu}_l},\\
  s_-(\wt{\nu}_ {j,l})=&~\sum_{j=1}^{2}\sum_{l=2}^{1}\FR{-2^{+\ii\wt{{\nu_j}}}\pi^{5/2}}{256\Gamma(1-\ii\wt{{\nu_j}})\sinh(\pi\wt{{\nu_j}})\big[\sinh(\pi\wt{{\nu_j}}/2)-
  \ii\cosh(\pi\wt{{\nu_j}}/2)\big]}\int_0^\infty\di z\,I_+^2(z)z^{-5/2-\ii\wt{\nu}_l}.
\end{align}
In the squeezed limit to show the periodic behaviour of  $S(k_1,k_2,k_3)$, we can rewrite it in the following form \cite{Chen:2017ryl}
\begin{align}\label{Spure}
  S_{pure}(k_1,k_2,k_3)=& P_\zeta^{-1/2} \left(\frac{\lambda_2}{H}\right)^3 \left ( \frac{\lambda_3}{H} \right )\Big(\FR{k_3}{k_1}\Big)^{1/2}\n\\
   \times& \sum_{j=1}^{2} \left \{
      s_1(\wt{{\nu_j}}) \sin\left [ \wt{{\nu_j}} \log\left ( \frac{k_3}{k_1}  \right ) \right ]
    + s_2(\wt{{\nu_j}}) \cos\left [ \wt{{\nu_j}} \log\left ( \frac{k_3}{k_1}  \right ) \right ]
  \right \},
\end{align}
where
\begin{align}
  s_1(\wt{{\nu_j}}) = \Re (s_+{\nu_j} - s_-{\nu_j}),\qquad
  s_2(\wt{{\nu_j}}) = \Im (s_+{\nu_j} + s_-{\nu_j}).
\end{align}
Here we wrote only the pure interacting parts and for mixed propagators one can repeat a same procedure as well.
 \begin{figure}[ht]
\centering
\includegraphics[scale=.750]{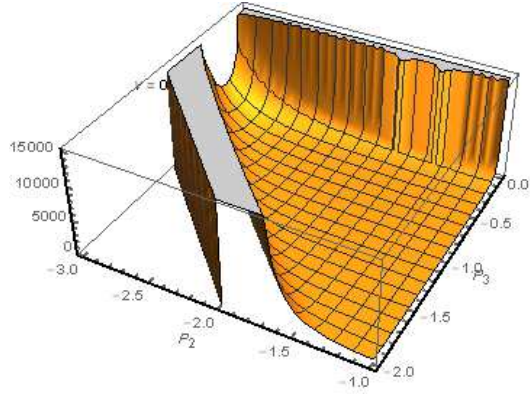}
\caption{{\it{This diagram shows the behavior of shape function expressed in eq.(\ref{Spure}) for the case $\nu=o$.  Here $P_2=\frac{k_2}{k_1}$ and $P_3=\frac{k_3}{k_1}$.}}}
\label{shape12}
\end{figure}
 \begin{figure}[ht]
\centering
\includegraphics[scale=.750]{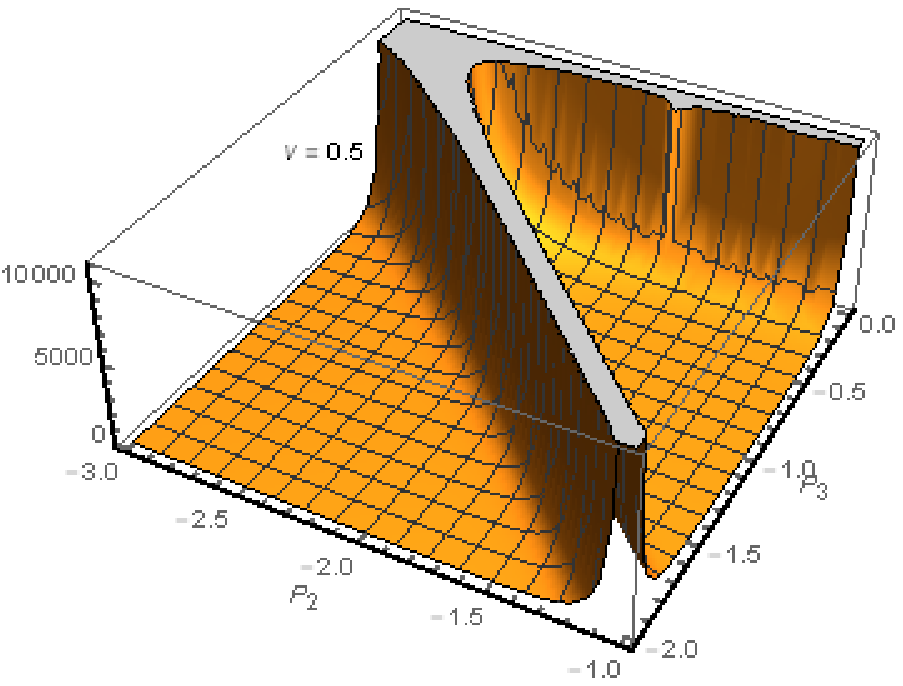}
\caption{{\it{This diagram expresses the behavior of shape function presented in eq.(\ref{Spure}) for the case $\nu=o.5$.}}}
\label{shape12a}
\end{figure}
 \begin{figure}[ht]
\centering
\includegraphics[scale=.750]{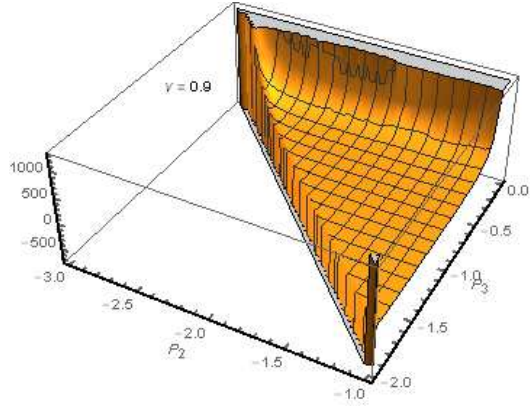}
\caption{{\it{This diagram indicates the behavior of shape function presented in eq.(\ref{Spure}) for the case $\nu=o.9$.}}}
\label{shape34b}
\end{figure}
 \begin{figure}[ht]
\centering
\includegraphics[scale=.750]{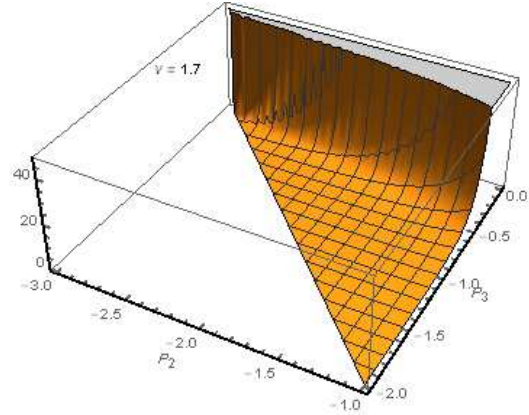}
\caption{{\it{This diagram expresses the behavior of shape function presented in eq.(\ref{Spure}) for the case $\nu=1.7$.}}}
\label{shape34}
\end{figure}
The figures \ref{shape12}-\ref{shape34} are related to the shape function appeared in Eq.(\ref{Spure}). In these shapes we consider $P_2=\frac{k_2}{k_1}$ and $P_3=\frac{k_3}{k_1}$. From these figures different limits based on different values of $\nu$ can be worked out.
\newpage

\section{Trispectrua}
\label{Trispectrum}

\begin{figure}[ht]
\centering
\includegraphics[scale=.80]{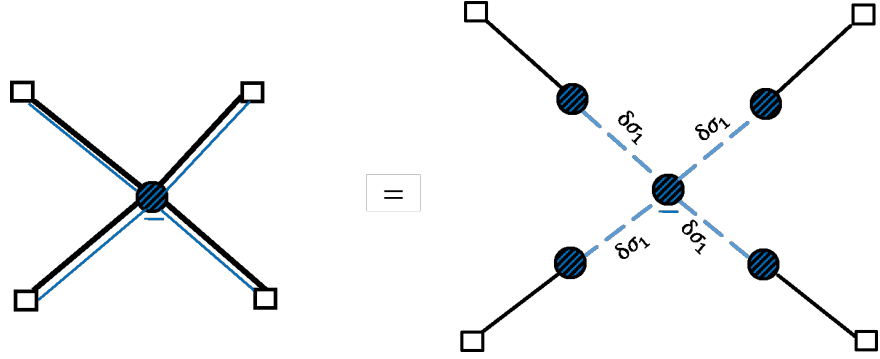}
\caption{{\it{This diagram shows the self interaction for quasi massive field $\delta\sigma_1$. The order of interacting for such a diagram is expressed as $\lam_{2j}^3\lam_{2l}^1\Lambda_{4j,l}^{}$}}}.
\label{fignsr1}
\end{figure}

\begin{figure}[ht]
\centering
\includegraphics[scale=.80]{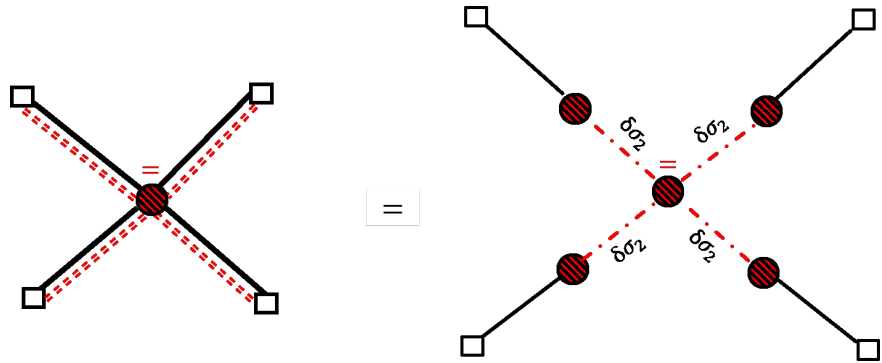}
\caption{{\it{This diagram expresses the self interaction for quasi massive field $\delta\sigma_2$. The order of interaction for such a diagram is  $\lam_{22}^4\lam_{42}^{}$}}}.
\label{fignsr2}
\end{figure}

\begin{figure}[ht]
\centering
\includegraphics[scale=.80]{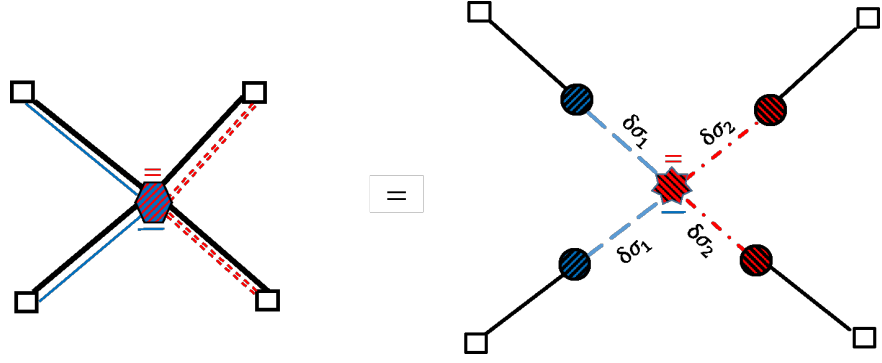}
\caption{{\it{This diagram shows the mixed self interaction for both quasi massive fields $\delta\sigma_j$ but when they have same portion in interaction. The order of interaction originated from terms with coefficient $\tilde{\Lambda}_{4j}\equiv6(\frac{\partial^2}{\partial^2 \si_1}(\frac{\partial^2}{\partial^2 \si_2}V))$ .}}}
\label{fignsr3}
\end{figure}

\begin{figure}[ht]
\centering
\includegraphics[scale=.80]{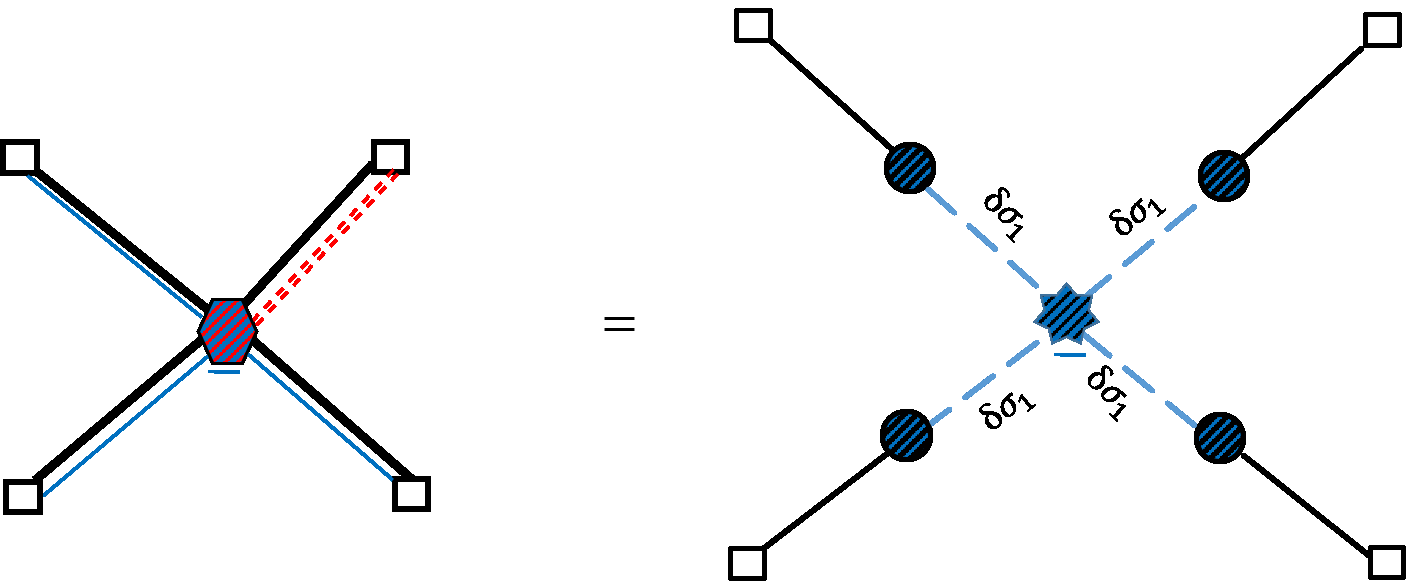}
\caption{{\it{This diagram indicates the mixed self interacting for both quasi massive fields $\delta\sigma_j$. The order of interacting originated from terms with coefficient ${\Lambda}_{41}\equiv4\frac{\partial}{\partial \si_2}(\frac{\partial^3}{\partial^3 \si_1}V)$.}}}
\label{fignsr4}
\end{figure}

\begin{figure}[ht]
\centering
\includegraphics[scale=.80]{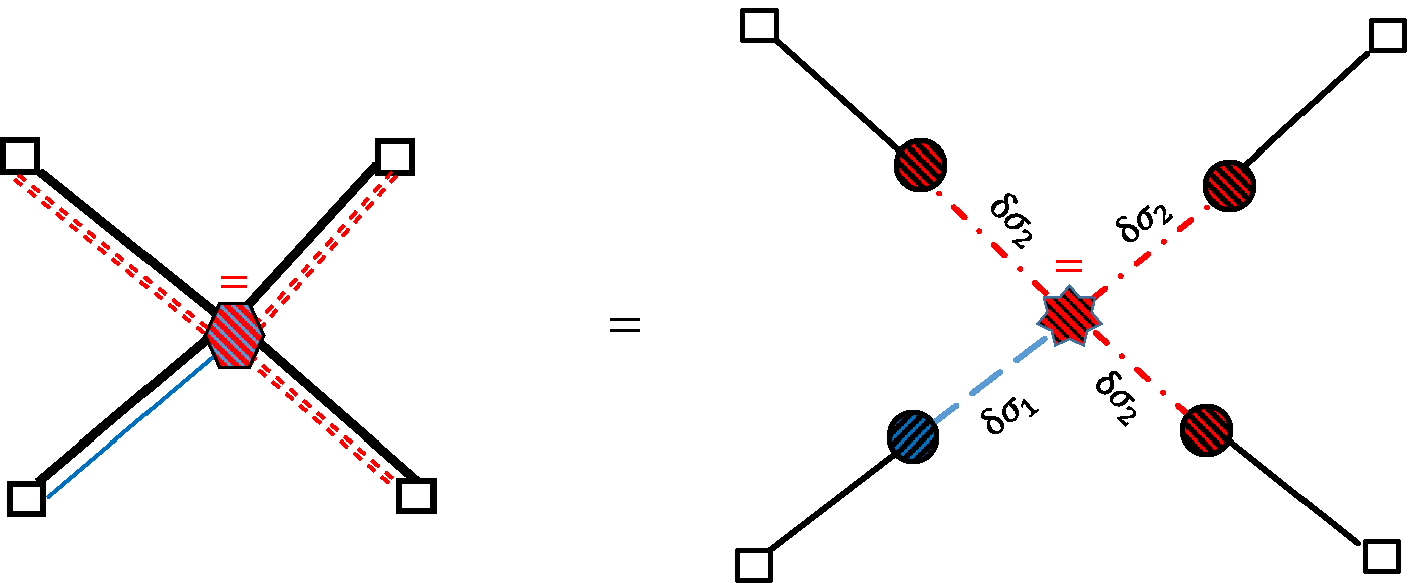}
\caption{{\it{This diagram shows the mixed self interaction for both quasi massive fields $\delta\sigma_j$. The order of interaction originated from terms with coefficient ${\Lambda}_{42}\equiv4\frac{\partial}{\partial \si_1}(\frac{\partial^3}{\partial^3 \si_2}V)$.}}}
\label{fignsr5}
\end{figure}

\begin{figure}[ht]
\centering
\includegraphics[scale=.80]{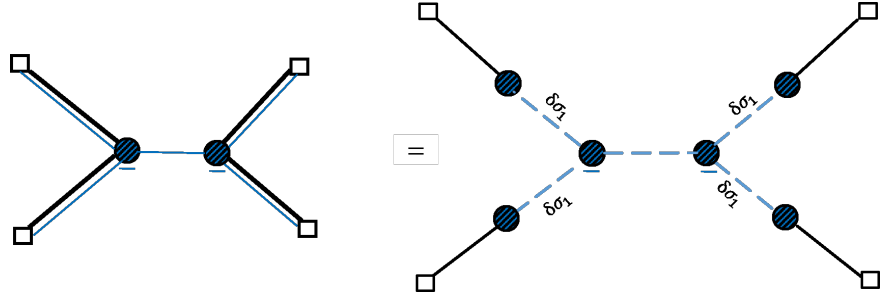}
\caption{{\it{This diagram shows the self interaction for quasi massive field $\delta\sigma_1$ but based on two cube portions.  The order of interaction originated from terms with coefficient $\lam_{2j}^3\lam_{2l}^1\Lambda_{4j,l}^{}$.}}}
\label{fignsr6}
\end{figure}

\begin{figure}[ht]
\centering
\includegraphics[scale=.80]{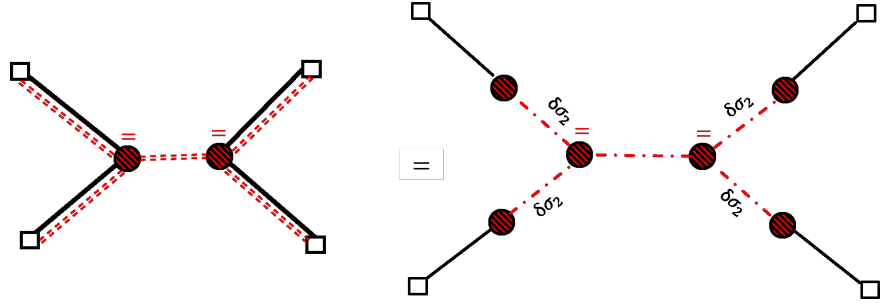}
\caption{{\it{This diagram expresses the mixed self interaction for  quasi massive field $\delta\sigma_2$ but for the case with two cube portions.  The order of interaction originated from terms with coefficient $\lam_{2j}^3\lam_{2l}^1\Lambda_{4j,l}^{}$.}}}
\label{fignsr7}
\end{figure}

\begin{figure}[ht]
\centering
\includegraphics[scale=.80]{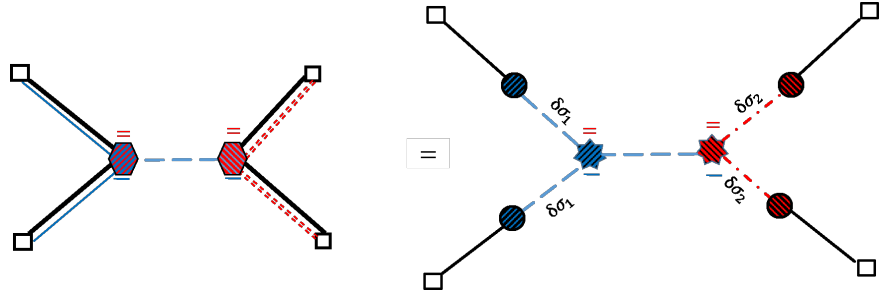}
\caption{{\it{This diagram indicates the mixed self interaction for  quasi massive field $\delta\sigma_2$ but based on same two cube portion.  The order of interaction risen from terms with coefficient $\lam_{2j}^2\lam_{2l}^2\tilde{\Lambda}_{4j,l}^{}$.}}}
\label{fignsr8}
\end{figure}

\begin{figure}[ht]
\centering
\includegraphics[scale=.80]{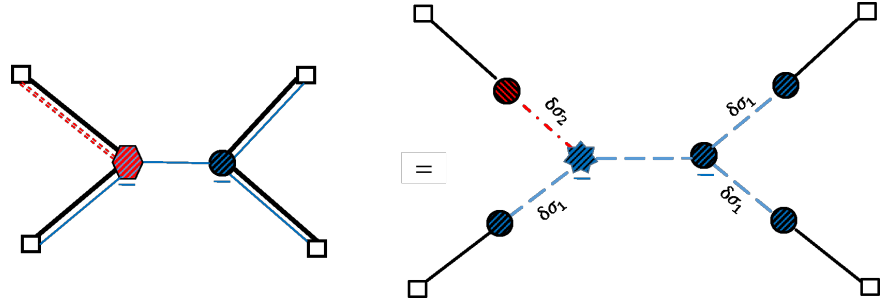}
\caption{{\it{This diagram shows the mixed self interacting for both quasi massive fields $\delta\sigma_j$ but based on two cube portions.  The order of interaction expressed by terms with coefficient $\lam_{2j}^3\lam_{2l}^1{\Lambda}_{3j,l}^{2}$.}}}
\label{fignsr9}
\end{figure}

\begin{figure}[ht]
\centering
\includegraphics[scale=.80]{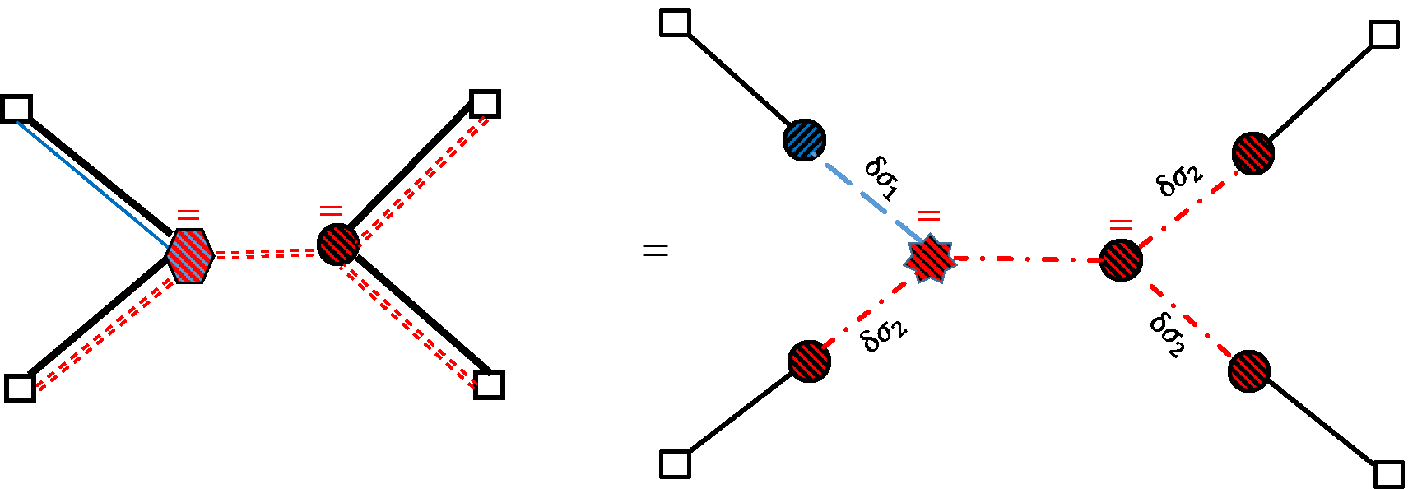}
\caption{{\it{It is same as Fig.\ref{fignsr9} but $\delta\sigma_2$ is dominant field.}}}
\label{fignsr10}
\end{figure}

The main purpose of this section is to calculate the leading four point function of the massless field in QSMF inflation. In other words the main idea for this calculation returns to calculating the trispectrum in multifid models and viewing the effects of such extra terms on amount of non-Gausianities. As we saw in bispectrum and powerspectrum modifications we expect that the mentioned extra effects  should be appeared for trispectrum explicitly. Following \cite{Chen:2018sce}, we can use, instead of $\de\phi$,  the curvature perturbation $\zeta=-H\de\phi/\dot\phi_0$ modes to parameterize the expectation values and receive the following relation for the shape function $T(\mb k_1,\mb k_2,\mb k_3,\mb k_4)$ as
\begin{equation}\label{4ptdef}
\la\zeta(\mb k_1)\zeta(\mb k_2)\zeta(\mb k_3)\zeta(\mb k_4)\ra'=(2\pi)^6 P^3_\zeta \frac{K^3}{(k_1k_2k_3k_4)^3}T(\mb k_1,\mb k_2,\mb k_3,\mb k_4),
\end{equation}
where $K=k_1+k_2+k_3+k_4$. To consider the leading portions of four-point function we have to consider both the self interactions  and the interaction between different types of semi-massive fields.  In the following we want to calculate explicitly the trispectrum for QSMF inflation by virtue of diagrammatic version of S-K formalism. In fact  we are going to find a suitable expression for the shape function \cite{Chen:2018sce}.
We should emphasise here that, two types of diagrams related to four-point function instead of one, comparing to the three-point function, must be considered.
By virtue of the Lagrangian expansion in Eq.(\ref{QBFL}), obviously seen that we have terms for leading four-point function diagrams. But this is not all the story and we should take care about portions of mixed interacting terms in different channels of four-point function. In other words, in QSMF inflation beside the self interacting terms the effects of interacting between quasi-massive fields have so important role. To show what we mean one can take look into the diagrams appeared in Figs.\ref{fignsr1}-\ref{fignsr9}. Let's explain in a bit more the details of these figures. In figures \ref{fignsr1}, \ref{fignsr2}, \ref{fignsr6} and {\ref{fignsr7}} we only see the pure self interacting portion of interactions of $\delta\sigma_{j}$. And the other remnant figures, i.e. Figs. \ref{fignsr3}, \ref{fignsr4}, \ref{fignsr5}, \ref{fignsr8}, \ref{fignsr9} and \ref{fignsr10}, refer to the mixed interacting parts related to  $\delta\sigma_{1}$ and $\delta\sigma_{2}$. So immediately one can realize that when we speak about the advantages of diagrammatic S-K formulation against the normal complicated integration method what we mean!.  One can use only one diagram and so one integration processing instead of 16 ones.  Following aforementioned sections, another interesting part of our investigation considering the trispectrum goes back to different channels of these diagrams namely $u$, $t$ and $s$ channels \cite{Chen:2017ryl,Chen:2018sce}.

The expression for this diagram is very simple, and it is similar to the case in bispectrum (\ref{3pt}). Now we are able to calculate the necessary steps to obtain the shape function for the quartic coupling. To do so we can write down the expectation values as
\begin{align}\label{4ptcontact00}
&~\la\de\phi(\tau,\mb k_1)\de\phi(\tau,\mb k_2)\de\phi(\tau,\mb k_3)\de\phi(\tau,\mb k_4)\ra_{\lam_4}'\n\\
=&~2\sum_{j=1}^{2}\lambda_{4j}
\text{Im} \int_{-\infty}^0 \frac{d\tau}{(-H\tau)^4}\mathcal G_{+j}(k_1;\tau)\mathcal G_{+j}(k_2;\tau)\mathcal G_{+j}(k_3;\tau) \mathcal G_{+j}(k_4;\tau)\n\\
=&~\sum_{j=1}^{2}\FR{\pi^4\lam_{2j}^4\lam_{4j}^{}}{2048k_1^3k_2^3k_3^3}\,\text{Im}\int_0^\infty\FR{\di z}{z^4}I_{+j}(\FR{k_1}{k_4}z)I_{+j}(\FR{k_2}{k_4}z)I_{+j}(\FR{k_3}{k_4}z)I_{+j}(z).
\end{align}
On the other hand, the $s$-channel diagram with two cubic-vertices can be written down as follows,
\begin{align}\label{raw4pt}
&\la\de\phi(\tau,\mb k_1)\de\phi(\tau,\mb k_2)\de\phi(\tau,\mb k_3)\de\phi(\tau,\mb k_4)\ra_{\lam_3^2,s}'\n\\
=&-\sum_{j=1}^{2}\lam_{2j}^4\sum_{p,n=\pm}\int_{-\infty}^0\FR{\di\tau_1\di\tau_2}{(H^2\tau_1\tau_2)^4} \mathcal{G}_{+j}(k_1;\tau_1)\mathcal{G}_{+j}(k_2;\tau_1)\mathcal{G}_{-j}(k_3;\tau_2)\mathcal{G}_{-j}(k_4;\tau_2)D_{\pm j}(k_s,\tau_1,\tau_2)\n\\
=&~\sum_{j=1}^{2}\frac{\pi^5 \lam_{2j}^4\lambda_{3j}^2}{8192H^2}\frac{k_s^3}{(k_1k_2k_3k_4)^3}e^{-\pi\,\textmd{Im}\,\nu} \,\textmd{Re}\,\bigg[\int_0^\infty\textmd{d}z \int_0^z \textmd{d}z' \Big(J_{12s}^{+-}(z)J_{34s}^{++}(z')+  J_{34s}^{+-}(z)  J_{12s}^{++}(z')\Big) \n\\
&- \int_0^\infty \textmd{d}z\int_0^\infty \textmd{d}z'J_{34s}^{+-}(z) J_{12s}^{-+}(z') \bigg].
\end{align}
where $k_s\equiv |\mb k_1+\mb k_2|$. Additionally, we have defined,
\begin{align}
  &J_{\varepsilon\zeta\eta}^{\pm +}(z)=\sum_{j=1}^2 z^{-5/2} I_{\pm j}\Big(\frac{k_\varepsilon}{k_\eta}z\Big)I_{\pm j}\Big(\FR{k_\zeta}{k_\eta}z\Big)\text{H}_{\nu_j}^{(1)}(z),
  &&J_{\varepsilon\zeta\eta}^{\pm -}(z)=z^{-5/2} I_{\pm j}\Big(\frac{k_\varepsilon}{k_\eta}z\Big)I_{\pm j}\Big(\FR{k_\zeta}{k_\eta}z\Big)\text{H}_{\nu_j^*}^{(2)}(z).
\end{align}
Now  by  substitution $(2\leftrightarrow 3, s\to t)$ and $(2\leftrightarrow 4, s\to u)$, respectively,  the expressions for $t$ and $u$-channels could be attained.
From the above results, we can find the shape function $T$ defined in (\ref{4ptdef}) as,
\begin{align}\label{4ptform}
 T_{pure}=&~\sum_{j=1}^2\FR{\pi^3}{2^{15}P_\zeta} \Big(\frac{\ka_1}{H}\Big)^4\bigg[\frac{4 \lambda_4}{\pi}t_{cj}+ \Big(\frac{\lambda_3}{H}\Big)^2\big(t_{sj}+t_{tj}+t_{uj}\big)\bigg],
\end{align}
where $t_{cj}$ and $t_{sj}$ are,
\begin{align}\label{4pttcoef}
t_{cj}=&~\,\textmd{Im}\,\sum_{j=1}^2\int_0^{\infty}\frac{\textmd{d}z}{z^4}I_{+j}\Big(\frac{k_1}{K} z\Big)I_{+j}\Big(\frac{k_2}{K} z\Big)I_{+j}\Big(\frac{k_3}{K} z\Big)I_{+j}\Big(\frac{k_4}{K} z\Big),\\
t_{sj}=&~\sum_{j=1}^2e^{-\pi \,\textmd{Im}\, \nu_j}\Big(\FR{k_s}{K}\Big)^3\,\textmd{Re}\,\bigg[\int_0^\infty\textmd{d}z \int_0^z \textmd{d}z'\Big(J_{12s}^{+-}(z)J_{34s}^{++}(z')+  J_{34s}^{+-}(z)  J_{12s}^{++}(z')\Big)\n\\
&~ -\int_0^\infty \textmd{d}z\int_0^\infty \textmd{d}z' J_{34s}^{+-}(z) J_{12s}^{-+}(z') \bigg],
\end{align}
and by virtue of the permutations rules we can write down  $t_t$ and $t_u$ as well   \cite{Chen:2017ryl}. Now to calculate the contribution of mixed propagators, and therefore related trispectum, one should repeat this procedure for coefficients of order ${\Lambda}_{41}\equiv4\frac{\partial}{\partial \si_2}(\frac{\partial^3}{\partial^3 \si_1}V)$, ${\Lambda}_{42}\equiv4\frac{\partial}{\partial \si_1}(\frac{\partial^3}{\partial^3 \si_2}V)$ and $\tilde{\Lambda}_{4j}\equiv6(\frac{\partial^2}{\partial^2 \si_1}(\frac{\partial^2}{\partial^2 \si_2}V))$. Beside this we have to consider the results of mixed interacting portions related to the bispectrum section \cite{Chen:2015lza,Chen:2016qce}. For example we present here some of mixed integrals as follows
\begin{align}\label{4ptcontact1}
&~\la\de\phi(\tau,\mb k_1)\de\phi(\tau,\mb k_2)\de\phi(\tau,\mb k_3)\de\phi(\tau,\mb k_4)\ra_{\Lambda_{4j,l}}'\n\\
=&~2\sum_{j=1}^{2}\sum_{l=2}^{1}\Lambda_{4j,l}
\text{Im} \int_{-\infty}^0 \frac{d\tau}{(-H\tau)^4}\mathcal G_{+j}(k_1;\tau)\mathcal G_{+j}(k_2;\tau)\mathcal G_{+j}(k_3;\tau) \mathcal G_{+l}(k_4;\tau)\n\\
=&~\sum_{j=1}^{2}\sum_{l=2}^{1}\FR{\pi^4\lam_{2j}^3\lam_{2l}^1\Lambda_{4j,l}^{}}{2048k_1^3k_2^3k_3^3}\,\text{Im}\int_0^\infty\FR{\di z}{z^4}I_{+j}(\FR{k_1}{k_4}z)I_{+j}(\FR{k_2}{k_4}z)I_{+j}(\FR{k_3}{k_4}z)I_{+l}(z)~,
\end{align}

\begin{align}\label{4ptcontact2}
&~\la\de\phi(\tau,\mb k_1)\de\phi(\tau,\mb k_2)\de\phi(\tau,\mb k_3)\de\phi(\tau,\mb k_4)\ra_{\tilde{\Lambda}_{4j,l}}'\n\\
=&~2\sum_{j=1}^{2}\sum_{l=2}^{1}\tilde{\Lambda}_{4j,l}
\text{Im} \int_{-\infty}^0 \frac{d\tau}{(-H\tau)^4}\mathcal G_{+j}(k_1;\tau)\mathcal G_{+j}(k_2;\tau)\mathcal G_{+l}(k_3;\tau) \mathcal G_{+l}(k_4;\tau)\n\\
=&~\sum_{j=1}^{2}\sum_{l=2}^{1}\FR{\pi^4\lam_{2j}^2\lam_{2l}^2\tilde{\Lambda}_{4j,l}^{}}{2048k_1^3k_2^3k_3^3}\,\text{Im}\int_0^\infty\FR{\di z}{z^4}I_{+j}(\FR{k_1}{k_4}z)I_{+j}(\FR{k_2}{k_4}z)I_{+l}(\FR{k_3}{k_4}z)I_{+l}(z)~,
\end{align}
and
\begin{align}\label{raw4pt1}
&\la\de\phi(\tau,\mb k_1)\de\phi(\tau,\mb k_2)\de\phi(\tau,\mb k_3)\de\phi(\tau,\mb k_4)\ra_{s}'\n\\
=&-\sum_{j=1}^{2}\sum_{l=2}^{1}\sum_{p,n=\pm}\lam_{2j}^3\lam_{2l}^1{\Lambda}_{3j,l}^{2}\n\\
&~\times\int_{-\infty}^0\FR{\di\tau_1\di\tau_2}{(H^2\tau_1\tau_2)^4} \mathcal{G}_{+j}(k_1;\tau_1)\mathcal{G}_{+j}(k_2;\tau_1)\mathcal{G}_{-j}(k_3;\tau_2)\mathcal{G}_{-j}(k_4;\tau_2)D_{\pm j}(k_s,\tau_1,\tau_2).
\end{align}
In the squeezed and collapsed limits the oscillatory signals again, like in the bispectrum section can begenerated.
For instance when we consider the $k_1/K\rightarrow 0$ for triangle limit in self interactions the equation \ref{4ptform} reduces to
\bge
\label{TriClock}
  T_{pure}=\sum_{j=1}^{2}\FR{\pi^3c_\phi^3 }{2^{15}P_\zeta} \Big(\frac{\ka_1}{H}\Big)^4\Big(\FR{k_1}{K}\Big)^{3/2}\Big[t_R\sin\Big(\wt\nu_j\log\FR{k_1}{K}\Big)+t_I\cos\Big(\wt\nu_j\log\FR{k_1}{K}\Big)+\cdots\Big],
\ede
and $t_R=\text{Re}(t_+ + t_-)$ and $t_I=\text{Im}(t_- - t_+)$.  Where
\begin{align}
  t_\pm =&~\FR{2^{1\mp\nu_j}\sin(\frac{\pi}{4}\pm\frac{\pi\nu_j}{2})C_\nu}{ \Gamma (1\pm\nu_j)\sin (\pm\pi\nu_j)}\bigg[-\FR{4\ii\lam_4}{\pi}\mathcal{I_j}_c^\pm+\Big(\frac{\lambda_3}{H}\Big)^2\big(\mathcal{I_j}_s^\pm+\mathcal{I_j}_t^\pm+\mathcal{I_j}_u^\pm\big)\bigg] ,
\end{align}
these integrals based on different channels are expressed as follows
\begin{align}
\mathcal{I_j}_c^\pm=&~ \sum_{j=1}^2\int_0^\infty\di z\,I_{+j}\Big(\frac{k_2}{K} z\Big)I_{+j}\Big(\frac{k_3}{K} z\Big)I_{+j}\Big(\frac{k_4}{K} z\Big)z^{-5/2\pm \nu_j}\\
\mathcal{I_j}_s^\pm=&~ \sum_{j=1}^2\Big(\FR{k_2}{K}\Big)^{3/2\mp \nu_j}e^{-\pi\,\text{Im}\,\nu_j}\int_0^\infty\textmd{d}z \int_0^z \textmd{d}z'\Big[z^{-1\pm \nu_j}J_{342}^{++}(z')I_{+j}(z)\text{H}_{\nu_j^*}^{(2)}(z) \n\\
&~+ z'^{-1\pm \nu_j}J_{342}^{+-}(z)I_{+j}(z')\text{H}_{\nu_j}^{(1)}(z')\Big]\n\\
 &~-\int_0^\infty \textmd{d}z\int_0^\infty z'^{-1\pm \nu_j}J_{342}^{+-}(z)I_{-j}(z')\text{H}_{\nu_j}^{(1)}(z').
\end{align}
To calculate Eq.(\ref{TriClock}) we should consider $\nu_j$ purely imaginary quantity \cite{Chen:2018sce}. For the mixed interaction, we can repeat the same procedure.

\section{Conclusions}\label{seccon}
The aim of this study was to calculate the leading order of bispectrum and trispectrum in an extended version of the quasi-single-field model
of inflation namely QSMF inflation. One expects that observable non-Gaussianity can be generated in multiple field models which can be tested observationally. Probing non-Gaussianity can bring models of multiple fields inflation in contact to observation.  Following \cite{Emami:2013lma} and \cite{Chen:2017ryl,Chen:2018sce}  we explicitly have calculated different expectation values for QSMF model namely power spectrum , bispectrum  and trispectrum.
In point of fact, our results showed that the  S-K diagrammatic formalism has some advantages compared to the usual in-in formalism. For instance, it dramatically reduces the number of calculations
and complication of working out the results.  The more interesting but complicated part of this calculation, in comparison to the quasi-single-filed, was appearing some extra terms beside the self-interaction parts, because of introducing an extra quasi-massive field.  In fact, in the quasi-single-field inflation most important terms are whose that merely deal with self-interaction and other remnant terms can be eliminated. In QSMF the story is completely different; what we mean goes back to the appearance of some extra mixed interacting terms due to the interaction between both quasi-massive fields and they played an important role in our calculations. As a consequence, it was expected that the amount of non-Gaussianity should show an increase in amount. So by virtue of  S-K diagrammatic rules, we tried out to check the accuracy of this claim in more details. To do so, at first we had to add some new diagrammatic rules to recognize the difference between interacting terms due to different presented fields. By looking at Figs. \ref{PSI3},  \ref{PSI3Aa} and \ref{Theta1} it will be realized that the amount of power spectrum in QSMF explicitly showed an increase. Additionally, if one makes a comparison between QSMF and quasi-single-field models immediately observes that the non-Gaussianities have undergone countable changes. For instance, if one considers the equations of shape function in  \cite{Chen:2017ryl,Chen:2018sce} and comparing with the corresponding equations in this work one finds out that the amounts of non-Gaussianity have a precious increasing in modules. Besides this, the prediction of quantum clocks could be concluded again. Consequently, we emphasize again that diagrammatic method dramatically decreased the amounts of perplexing calculations in term of calculation of the bispectrum and especially trispectrum. And finally, it was obviously seen that our results have good agreement with previous literature.

\section*{Acknowledgment}\label{secconAck}
The author would like to thank Hassan Firouzjahi for his collaboration during all stages of this work. He also is grateful the Institute for Research in Fundamental Sciences (IPM) for their hospitality during his visit and support in part.





\end{document}